\documentclass[aps,prl,twocolumn,superscriptaddress,notitlepage]{revtex4-2}

\usepackage{amsmath,amssymb,bm}
\usepackage{graphicx}
\usepackage{booktabs}
\usepackage{physics}           
\usepackage[toc]{appendix}
\usepackage{enumitem}
\usepackage[normalem]{ulem}
\usepackage[dvipsnames,usenames]{xcolor}
\colorlet{Green}{green}
\colorlet{Blue}{blue}
\colorlet{Red}{red}
\colorlet{Purple}{purple}

\newcommand{\PRLsection}[1]{%
  \par\vspace{1mm}%
  \noindent\textcolor{black}{\textit{#1.—}}\ %
}
\usepackage[
  colorlinks=true,
  linkcolor=blue,
  citecolor=blue,
  urlcolor=blue,
  breaklinks=true,
  bookmarks=true
]{hyperref}

\begin{document}
\title{Second-Harmonic Bulk Response of Fractional Quantum Hall Fluids}
\author{J.~Luke Pimlott} 
\affiliation{Department of Physics, University of Bath, Claverton Down, Bath BA2 7AY, United Kingdom}

\author{Pablo Reiser} 
\affiliation{Department of Physics, University of Bath, Claverton Down, Bath BA2 7AY, United Kingdom}

\author{T.~Hans~Hansson}
\affiliation{Fysikum, Stockholm University, 106 91 Stockholm, Sweden}

\author{Habib Rostami}
\email{hr745@bath.ac.uk}
\affiliation{Department of Physics, University of Bath, Claverton Down, Bath BA2 7AY, United Kingdom}

\date{\today}
 
\begin{abstract}
We develop a bulk description for the second-harmonic Hall response of
fractional quantum Hall fluids with broken inversion and rotational
symmetries, distinct from edge theories based on nonlinear edge-mode
dispersion. By projecting a tilted massive Dirac model onto a single
low-energy band, we construct a composite-boson Chern--Simons theory
for the lowest Landau level that retains projection-induced nonlinear
electromagnetic couplings. The nonlinear Hall resistivity vanishes in
the dc limit, while its finite-frequency response retains the
characteristic filling-factor dependence
$\rho^{(2)}_{TLL}\propto\nu^{-2}$, where $\nu$ is the Landau-level
filling factor. At low frequency, its real and imaginary parts scale as
$(\omega/\omega_c)^2$ and $\omega/\omega_c$, respectively, where
$\omega_c$ is the cyclotron frequency, and the response is strongly
anisotropic with respect to the crystalline tilt axis. Notably, the Kohn-mode resonances
present in the second-order conductivity cancel in the nonlinear Hall
resistivity. These results establish finite-frequency nonlinear
transport as a bulk probe of fractional quantum Hall fluids without
requiring nonlinear edge-mode dispersion.
\end{abstract}

\maketitle

\PRLsection{Introduction}
At low temperature, and in a large magnetic field, a clean two-dimensional electron gas will form a liquid state where the Hall conductance is quantized in integer units of $ e^2/h $~\cite{Klitzing1980} to a precision of $10^{-9}$. Originally, this integer quantum Hall effect (QHE) was explained using arguments based on gauge invariance~\cite{Laughlin1981}, and later by analysis of edge modes~\cite{Halperin1982}, and topological invariants~\cite{TKNN1982}. 
Broken time-reversal symmetry in magnetic materials or Chern insulators can lead to both classical~\cite{Nagaosa2010} and quantum~\cite{Chang2023} anomalous Hall effects~\cite{Karplus1954, Haldane1988}, and this has been experimentally observed in various systems, including magnetic topological insulators~\cite{Deng2020} and the orbital ferromagnets with Chern insulator moiré heterostructures \cite{Chang2023}.

Recent advances have revealed a rich landscape of {\em nonlinear Hall effects} \cite{Du2021}. In noncentrosymmetric but time-reversal
symmetric (TRS) systems, a transverse current can be rectified at second order in an oscillating electric field, even in the absence of an external magnetic field \cite{Moore2010,Sodemann2015}. The classical nonlinear Hall effect arises in metallic systems and is governed by quantum geometric properties—most notably the {\em Berry curvature dipole} \cite{Sodemann2015}, or, in PT-symmetric systems where the Berry curvature vanishes, the {\em quantum metric dipole} \cite{Gao2014}. These effects have been experimentally observed in low-symmetry materials such as WTe$_2$ \cite{Ma2019,Kang2019}, and more recently in PT-symmetric magnetic topological insulators \cite{Gao2023}.
\begin{figure}
    \centering
    \includegraphics[width=0.95\linewidth]{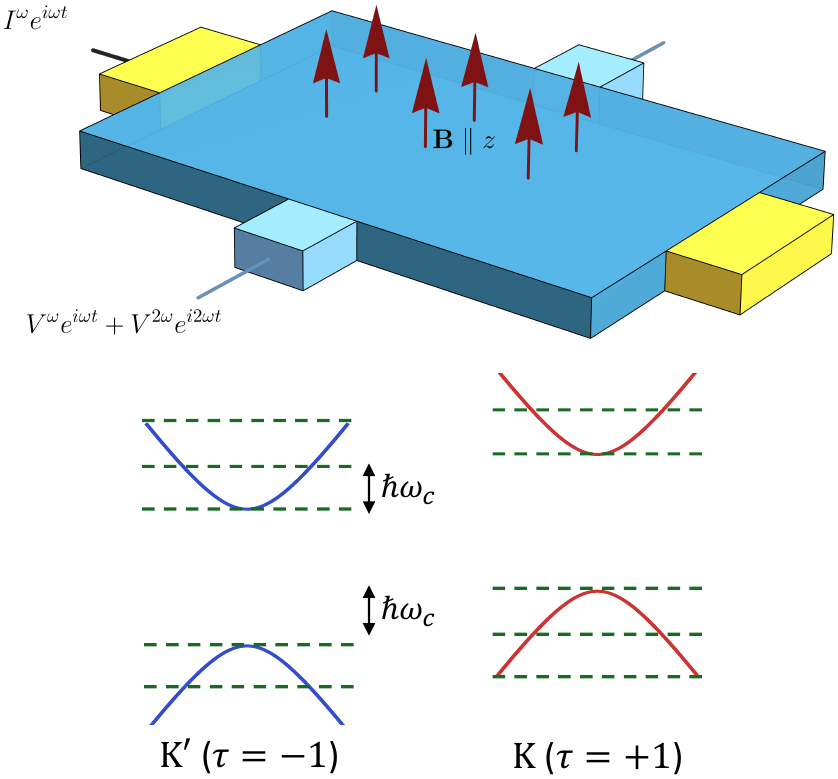}
\caption{
Top: Hall-bar geometry in an out-of-plane magnetic field
$\mathbf{B}\parallel\hat{\mathbf z}$, driven by an oscillating current
$I^\omega e^{i\omega t}$. The transverse voltage contains first- and
second-harmonic components,
$V^\omega e^{i\omega t}+V^{2\omega}e^{2i\omega t}$.
Bottom: Schematic low-energy Landau-level spectrum near the
$K'$ ($\tau=-1$) and $K$ ($\tau=+1$) valleys of an untilted massive
Dirac model. Green dashed lines denote the Landau levels. The
$n\geq1$ levels are valley degenerate, whereas the $n=0$ Landau level
is valley split and occurs at opposite band edges in the two valleys.
}
    \label{fig:sketch}
\end{figure}
Although nonlinear Hall effects have been studied extensively in TRS  systems, their TRS broken counterparts are attracting growing interest~\cite{He2019,Rikken2005}. In magnetic insulators lacking inversion and rotational symmetries, a second-order Hall current, $j^{\rm NL}_{\rm H}\propto E^2$, can arise, with strained twisted bilayer graphene providing a candidate platform~\cite{Kaplan2023}. Unlike the linear Hall response, this correction is not fixed by the first Chern number, and may remain finite at $C=0$. It originates from virtual interband transitions analogous to subgap photocurrents in Haldane Chern insulators~\cite{Kaplan2020}. These results raise the broader question of the robustness of Hall quantization beyond linear response, long studied in the context of current-induced QHE breakdown~\cite{Streda1986,Nachtwei1999}.

Recent experiments and theory~\cite{He2024,Isobe2024} have revealed a third-order Hall current, $j^{\rm NL}_{\rm H}\propto E^3$, in graphene's quantum Hall insulating regime under a strong perpendicular magnetic field and low-frequency electric drive.  
Although frequency dependent, the third-harmonic Hall voltage exhibits plateaus across the $\nu=\pm2$ quantum Hall states as the gate voltage is varied~\cite{He2024}. An edge mechanism based on interactions between counterpropagating
chiral modes with nonlinear dispersions has been proposed
\cite{Isobe2024}. A complementary classical hydrodynamic mechanism for
nonlinear quantum Hall response was recently proposed
\cite{Isobe2026Hydrodynamic}. Analogous finite-frequency plateaus occur
in graphene's Faraday and Kerr rotations and reflect the optical Hall
conductivity \cite{Shimano2013}. However, a microscopic bulk quantum
theory of nonlinear QHE plateaus remains lacking. Here, we develop such a theory up to order $E^2$. 

In this Letter, we address this gap by providing a theory for this response based on a perturbative expansion in an effective Ginzburg-Landau-Chern-Simons (GLCS) theory~\cite{ZHK,ZHANG1992,Hansson2017,ferraz2020strongly}.
We begin with a tilted Dirac insulator~\cite{Sodemann2015,rostami2020} describing the low-energy bands of WTe$_2$~\cite{Ma2019}. Using a field-theoretical approach~\cite{Son2013,ZHVQ}, we project the two-band model onto a single low-energy band while retaining the projection-induced nonlinear light--matter coupling.  
While such corrections are generally allowed by symmetry, our method predicts their magnitude, and is easily generalizable to multiband systems. These corrections are then incorporated in the GLCS theory to allow for a perturbative calculation of nonlinear Hall response. Finally we discuss how our results could be tested in experiments, and provide an outlook on the expected behaviour of higher-order nonlinear quantum Hall currents in generic multiband systems.  The measurement geometry and valley-resolved Landau-level structure
considered below are illustrated in Fig.~\ref{fig:sketch}.
 
\PRLsection{Nonlinear quantum Hall resistivity}
For a low-frequency current-density drive
$\boldsymbol j(t)=\mathrm{Re}[\boldsymbol j^\omega e^{i\omega t}]$,
we calculate the response
$\boldsymbol E(t)=\mathrm{Re}[\boldsymbol E^\omega e^{i\omega t}
+\boldsymbol E^{2\omega}e^{2i\omega t}]$.
We first compute the conductivity tensors and then invert the response
order by order to obtain the nonlinear resistivity. In the basis
$\hat{\boldsymbol e}_{L}\parallel\boldsymbol j^\omega$ and
$\hat{\boldsymbol e}_{T}=\hat{\boldsymbol z}\times\hat{\boldsymbol e}_{L}$,
the conductivity response reads
\begin{equation}
j_a^\omega
=
\sigma_{ab}(\omega)E_b^\omega
+
\sum_{\omega'}
\sigma^{(2)}_{abc}(\omega;\omega',\omega-\omega')
E_b^{\omega'}E_c^{\omega-\omega'} ,
\end{equation}
where $a,b,c\in\{L,T\}$. We assume the insulating regime,
$\omega,\omega_c\ll\Delta/\hbar$, where real interband absorption is
suppressed and the second-order rectified transport response vanishes.
Inverting this relation order-by-order gives~\cite{Principi2019}
\begin{equation}
\rho_{TLL}^{(2)}(2\omega)
=
-\rho_{Ta}(2\omega)
\sigma_{abc}^{(2)}(2\omega)
\rho_{bL}(\omega)
\rho_{cL}(\omega),
\label{eq:rho_TLL}
\end{equation}
with $\rho_{ab}=\sigma^{-1}_{ab}$ and
$\sigma_{abc}^{(2)}(2\omega)\equiv\sigma^{(2)}_{abc}(2\omega;\omega,\omega)$.
The nonlinear transverse resistivity is not determined
only by $\sigma_{TLL}^{(2)}$, but also by the linear Hall resistivity
through the three factors of $\rho_{ab}$. In an isotropic system, the Hall resistivity, $\rho_{H}\equiv\rho_{TL}=-\rho_{LT}$, is odd with respect to magnetic field, but in the second-order regime the situation is more subtle because the nonlinear resistivity depends on both the second-order conductivity and the linear resistivity tensor. 
Consequently, the transverse second-order resistivity need not be odd in $\mathbf B$ and can remain finite in noncentrosymmetric TRS metals through Berry-curvature and quantum-metric dipoles. These Fermi-surface contributions vanish in a gapped quantum Hall state, allowing the field-odd response to isolate the nonlinear quantum Hall resistivity. 
For a TRS system, $\rho_{H}=0$, and
Eq.~\eqref{eq:rho_TLL} reduces to
\begin{equation}
\rho_{TLL}^{(2)}(2\omega)
=
-\rho_{TT}(2\omega)
\sigma_{TLL}^{(2)}(2\omega)
\rho_{LL}^2(\omega)\,,
\end{equation}
which is the classical nonlinear Hall effect. By contrast, when a magnetic
field breaks time-reversal symmetry, $\rho_{H}\neq0$, and
$\rho_{TLL}^{(2)}$ contains both the intrinsic field-odd part of
$\sigma_{abc}^{(2)}$ and mixing with the ordinary linear Hall response.

In the Landau-quantized insulating regime considered below, the semiclassical Berry curvature dipole physics is frozen out, and the nonlinear response is instead controlled by the projected quantum geometry and collective dynamics of the occupied Landau levels. 
In the ideal dc QH limit, the linear resistivity is purely Hall-like:
$\rho_{LL}(\omega),\rho_{TT}(\omega)
=(h/\nu e^2)\mathcal{O}(\omega^n)$, with $n\geq1$, while
$\rho_{TL}(0)=-\rho_{LT}(0)\simeq h/(\nu e^2)$.
For the model considered below, we find
$\sigma^{(2)}_{abc}(2\omega)=\nu(e^3/h)\mathcal{O}(\omega^m)$,
with $m\geq1$. Equation~\eqref{eq:rho_TLL} therefore gives the
characteristic scaling $\rho^{(2)}_{TLL}\propto\nu^{-2}$. Its low-frequency form is
\begin{equation}
\rho_{TLL}^{(2)}(2\omega)
\simeq
\frac{h^2}{\nu^2 e^3}\,
\left[
\left(\frac{\omega}{\omega_c} \right)^2f(\theta)+i \left(\frac{\omega}{\omega_c} \right) g(\theta)
\right],
\label{eq:rho_qh_LT}
\end{equation}
where $\omega_c$ is the cyclotron frequency and the dimensionful functions \(f(\theta)\) and \(g(\theta)\) encode the direction of the current drive relative to the crystalline tilt axis set by angle $\theta$.
Although the nonlinear Hall resistivity vanishes in the strict dc limit, at finite  $\omega$ it retains the characteristic \(1/\nu^2\)  scaling with a non-universal prefactor. To leading order in frequency, $\omega\ll\omega_c$, the linear
resistivity is Hall dominated, and Eq.~\eqref{eq:rho_TLL} reduces to $\rho_{TLL}^{(2)}(2\omega)
\sim
-\rho_H^3\sigma_{LTT}^{(2)}(2\omega)$. 
On the other hand, components such as
$\sigma_{LLL}^{(2)}$ and $\sigma_{TTT}^{(2)}$, which are not themselves
Hall-like, also contribute when the longitudinal resistivities
$\rho_{LL}(\omega)$ and $\rho_{TT}(\omega)$ are finite, but are
subleading in the Hall-dominated regime.

\PRLsection{Effective model for Dirac electrons in a magnetic field}
A generic two-band system is described by the Dirac Lagrangian
\begin{align} \label{sch}
{\cal L} =  \Psi^\dagger (i\hbar\partial_t +eA_0 - \hat H(\bm \pi))  \Psi \,,
\qquad
\Psi =
\begin{pmatrix}
    \psi \\
    \chi
\end{pmatrix}\, ,
\end{align}
where $\psi$ and $\chi$ are the spinor components corresponding
to the conduction and valence bands, respectively.
For a TRS system with broken inversion and rotational symmetries, we take the Hamiltonian, 
\begin{align}
    \hat H(\boldsymbol{\pi}) = d_0(\boldsymbol{\pi}) \hat I + \boldsymbol{d}(\boldsymbol{\pi}) \cdot \hat{\boldsymbol{\sigma}} \,,
\end{align}
where $\hat I$ is the identity matrix, $\hat{\boldsymbol{\sigma}} = (\hat{\sigma}_x,\hat{\sigma}_y,\hat{\sigma}_z)$ are the Pauli matrices, and
$\boldsymbol{\pi} = \boldsymbol{p} + e \boldsymbol{A}$, where  
$[\pi_x,\pi_y] = - i\hbar e B$, 
with $e>0$, and $B$ the external magnetic field. 
For a finite gap $2\Delta$, we write $d_z({\bm\pi}) = \Delta + \eta({\bm\pi})$ with $\eta({\bm 0})=0$. To project onto a single band, we first make the transformation $\Psi \rightarrow e^{i\Delta t/\hbar}\Psi$, which shifts the conduction band energy to zero and the valence band to $-2\Delta$, and then integrate out the “heavy” valence band component $\chi$, and expand the result in inverse powers of $\Delta $ to get
\begin{equation} \label{effact}
S_{\text{eff}}[\psi^\dagger,\psi] = \int d^3 {\bm x} {\cal L}_{\rm eff} = -i\hbar \ln \int \mathcal{D}[\chi^\dagger,\chi]\, e^{iS/\hbar},
\end{equation}
where $S = \int d^3 \, {\bm x} {\cal L}$, and 
\begin{align}\label{EffL}
   & {\cal L}_{\rm eff} =\psi^\dagger\bigg
    \{i\hbar\partial_t +eA_0 - d_0({\bm \pi})-\eta({\bm \pi}) 
    -d_{-}({\bm \pi}) \, {\cal G} \, 
    d_{+}({\bm \pi})\bigg\}\psi
\end{align}
where ${\cal G}({\bm \pi},i\hbar\partial_t) = [i\hbar\partial_t  +eA_0 +2\Delta  - d_0({\bm \pi})+\eta({\bm \pi})]^{-1}$ and $d_{\pm}({\bm \pi})=d_x({\bm \pi}) \pm id_y ({\bm \pi})$. 
For a sufficiently large band gap $\Delta$ we expand the resolvent as
\begin{equation}
{\cal G}({\bm \pi},i\hbar\partial_t)
\approx \frac{1}{2\Delta}
      -\frac{i\hbar\partial_t +eA_0 - d_0 + \eta}{(2\Delta)^2}
      +\dots,
\end{equation}
to get the effective action as a polynomial in the dynamical fields.
We then expand the ${\bm d}$‑vector
in powers of the small kinetic momentum ${\bm \pi}$ about the two valley points,
defined by ${\bm \pi}=0$, where the band gap is minimal. We absorb the first‑order time derivative by introducing a renormalisation field factor order‑by‑order in the band‑gap expansion, and  neglect higher‑order time derivatives, which is justified in the low‑frequency regime. 

We consider an inversion-broken massive Dirac model whose tilt breaks
rotational symmetry and permits a second-order Hall response,
\begin{equation}
(d_0,d_x,d_y,d_z)
=
v(\alpha\tau\pi_x,\tau\pi_x,\pi_y,\Delta/v).
\end{equation}
where \(\tau=\pm1\) labels the valleys, \(\alpha\) is the tilt parameter, \(v\) is the Dirac velocity, and \(2\Delta\) is the band gap. With this, the projection onto the low-energy band gives (see Supplemental Material (SM)~\cite{sm} for details)
\begin{align} \label{eq:effective_lagrangian}
&\hspace{-2mm}\mathcal{L}
=
\psi^\dagger
\Bigg[
\big(i\hbar\partial_t+eA_0\big)
-\frac{|{\bm \pi}|^2}{2m}
-\alpha\left(\tau+\frac{\hbar\omega_c}{4\Delta}\right)v\pi_x
 \\
&
-\tau\frac{e\hbar}{4(mv)^2}
\boldsymbol{E}\cdot(\boldsymbol{\pi}\times\hat{\boldsymbol z}) 
-\frac{e\hbar\alpha}{4(mv)^3}
\left(
E_x\frac{\{\pi_y,\pi_x\}}{2}
-
E_y\pi_x^2
\right)
\nonumber\\
&
+
\frac{|{\bm \pi}|^4}{8m^3v^{2}}
\Bigg]\psi 
-\tau
\left[
\frac{\hbar\omega_c}{2}
-
\frac{\alpha(e\hbar)^2}{8(mv)^3}E_yB
\right]\psi^\dagger\psi
+
\mathcal{O}\!\left(\frac{\alpha\pi^4}{m^3v^{2}}\right),\nonumber
\end{align}
where \(m=\Delta/v^2\), \(\omega_c=eB/m\), and \(A_0\) is the external scalar
potential.
The first two terms describe the massive low-energy band in a magnetic field,
while the term linear in \(\pi_x\) is the effective tilt, including its leading
field correction. The electric field coupling
\(\boldsymbol{E}\cdot(\boldsymbol{\pi}\times\hat{\boldsymbol z})\) is the
anomalous-position contribution generated by the Berry connection
$\boldsymbol{\mathcal A}
=
\frac{\tau\hbar}{4m^2v^2}
\boldsymbol{\pi}\times\hat{\boldsymbol z}$.
The tilt-dependent term quadratic in momentum is the leading projected operator
that contributes to the intrinsic second-order electromagnetic response. The
\(\boldsymbol{\pi}^4\) term is subleading for the low-frequency response
but must be retained when discussing ultraviolet convergence of loop
corrections.

The momentum-independent terms in the second square brackets encode valley-dependent magnetic and
magnetoelectric shifts. The first,
\(\tau\hbar\omega_c/2\), is the valley Zeeman energy generated by the orbital
magnetic moment associated with Berry curvature; it shifts the two valleys in
opposite directions and lifts their degeneracy by \(\hbar\omega_c\). Such
valley Zeeman shifts are well established in monolayer transition-metal
dichalcogenides, including MoSe$_2$ \cite{Li2014,MacNeill2015}, WSe$_2$
\cite{Srivastava2015,Aivazian2015}, and MoS$_2$/WS$_2$ at high fields
\cite{Stier2016}. Remote bands further generate particle--hole asymmetry in the
effective valley Zeeman coupling, making the splitting directly visible in
magneto-photoluminescence and polarization-resolved spectroscopy
\cite{Rostami_2015}. The second momentum-independent term, proportional to
\(\tau\alpha E_yB\), is a tilt-induced magnetoelectric contact term coupling the
electric field, magnetic field, and valley index.

We account for the valley Zeeman splitting by computing the response separately in each valley, with valley-dependent filling factors, and summing over \(\tau=\pm1\). The
\(\mathcal{O}(\alpha\pi^4/m^3v^2)\) operators are neglected for small tilt,
and the \(\alpha^0\pi^4\) correction is kept only where it is required for
ultraviolet regularization. 

\PRLsection{Composite-boson theory of nonlinear QHE}
We start from the generalized GLCS action,
\begin{equation}
S
=
\int d^3x\,\psi^\dagger
\bigl(
i\hbar\partial_t+eA_0-\mathcal H(\pi_i)
\bigr)\psi
-
\int d^3x\,V[\rho],
\label{eq:fermion_action}
\end{equation}
where $\rho=\psi^\dagger\psi$, and where the single-particle Hamiltonian in Eq. \eqref{eq:effective_lagrangian} is expanded as
\begin{equation}
\mathcal H(\pi_i)
=
\frac{\pi^2}{2m} - \frac{1}{2m v^{2}} \left(\frac{\pi^2}{2m}\right)^2 
+
(c_{i;\alpha}+c_{i;E})\pi_i
+
c_{ij;E}\pi_i\pi_j 
+\cdots \, .
\label{eq:H_expand}
\end{equation}
Here $m$ is the effective mass, $c_{i;\alpha}$ contributes only to the
linear response, and $c_{i;E}$ and $c_{ij;E}$ encode the
electric-field-dependent corrections generated by band projection.
For the tilted Dirac model, we have 
\begin{align}
c_{i;\alpha}
&= \alpha \left(\tau+\frac{\hbar\omega_c}{4\Delta}\right) v\,\delta_{ix},\quad
c_{i;E}=
\frac{e\tau\hbar}{4m^2v^2} \epsilon^{ji}E_j,
\nonumber\\
c_{ij;E}
&=\frac{-e\hbar\alpha}{4m^3v^3}
\left[\delta_{ix}\delta_{jx}E_y
-
\frac{\delta_{iy}\delta_{jx}+\delta_{ix}\delta_{jy}}{2}E_x
\right].
\label{eq:cij_tilted}
\end{align} 

The transformation to composite bosons is via a singular gauge transformation that attaches an odd number of ``statistical" fluxes to the electrons by coupling to a statistical gauge field $a_\mu$ with a Chern-Simons action. The full action is,
\begin{align}
S
&=
\int d^3x\,
\phi^\dagger
\left[
i\hbar\partial_t-a_0+eA_0
-\mathcal H(-i\hbar\nabla+e\mathbf A-\mathbf a)
\right]\phi
\nonumber\\
&\quad
+
\frac{\nu}{4\pi\hbar}
\int d^3x\,
\epsilon^{\mu\nu\lambda}
a_\mu\partial_\nu a_\lambda  
-
\int d^3x\,V[\rho] \, ,
\label{eq:cb_action}
\end{align}
where variation with respect to $a_0$ gives
$\epsilon^{ij}\partial_i a_j=(2\pi\hbar/\nu)\rho$, with
$\nu=(2p+1)^{-1}$, attaching $2p+1$ statistical flux quanta to each
particle. Here $p=0$ corresponds to the $\nu=1$ IQH state, while
$p\geq1$ gives the Laughlin FQH sequence, and $V[\rho]$ is the
interaction energy of the composite bosons with density
$\rho=|\phi|^2$.
Writing $\phi=\sqrt{\rho}\,e^{i\theta}$ and absorbing the phase into the gauge field, $a_\mu\to a_\mu+\hbar\partial_\mu\theta$, we can expand about the mean-field state where $b = eB$ so the statistical gauge field cancels the external magnetic field, and we can choose $a_i=eA_i+\delta a_i$ where $\delta a_i$ describes the low-energy gauge fluctuation. The action (excluding the $\boldsymbol{\pi}^4$-term) then becomes
\begin{align}
S
&=
\int d^3x
\Bigg[
\frac{i\hbar}{2}\partial_0\rho
-
\rho\,\delta a_0
-
\frac{\left| \mathbf{Q}\right|^2}{2m\rho} 
+
(c_{i;\alpha}+c_{i;E})Q_i
\nonumber\\
&
-
\frac{c_{ij;E}Q_iQ_j}{\rho}
-
V[\rho]
\Bigg]
+
S_{\rm CS}[eA_\mu+\delta a_\mu],
\label{eq:hydro_action}
\end{align}
where \(Q_i\equiv \rho\,\delta a_i+(i\hbar/2)\partial_i\rho\), so that
\(|Q_i|^2=\rho^2(\delta a_i)^2+(\hbar^2/4)(\partial_i\rho)^2\), combining the gauge-field kinetic and quantum-pressure terms.
Expanding around the uniform saddle point,
\(\rho=\bar\rho+\delta\rho\), gives
\begin{equation}\label{eq:density_constraint}
\bar\rho
=
\frac{\nu}{2\pi\hbar}eB,
\qquad
\delta\rho
=
\frac{\nu}{2\pi\hbar}\delta b,
\qquad
\delta b=\epsilon^{ij}\partial_i\delta a_j \, ,
\end{equation}
and we take \(V[\rho]\simeq K \delta\rho^2/2 + {\cal O}(\delta \rho^{n>2})\), where \(K\) is the inverse compressibility.  
Using Eq.~\eqref{eq:density_constraint} to eliminate $\delta\rho$, the low-energy action can be written as 
\begin{align}
\label{eq:S_da}
S[\delta a, A] &=
\int d^3x
\bigg[
-\frac12\,\delta a_i [G^{-1}]^{ij}\delta a_j
 -\bar{\rho} c_{ij;E}\,\delta a_i\delta a_j
\nonumber\\ &+
\mathcal J_i\delta a_i
-
\mathcal L[A]
\bigg] \, .
\end{align}
The  Gaussian kernel is
\begin{align}
&[G^{-1}]^{ij}
=
\frac{\nu}{2\pi\hbar}
\left(
\omega_c\delta^{ij}
-
\partial_0\epsilon^{ij}
\right)
-
\frac{\nu^2 K}{4\pi^2 \hbar^2} \epsilon^{ki}\epsilon^{lj}
\partial_k\partial_l 
\nonumber\\&
+ \frac{\nu \hbar\omega_c}{8\pi (m v)^2}
\partial_i \partial_j
+ {\cal O}(\nabla^4)
\label{eq:Ginv}
\end{align}
where the third term originates from the retained quartic kinetic correction, and we neglect 
fourth and higher-order spatial derivative terms such as the quantum pressure.

The source coupled to $\delta a_i$ is written in the compact form 
\begin{align}
\mathcal J_i=
-\frac{\nu e}{2\pi\hbar}\epsilon^{ij}E_j
+
\bar\rho (c_{i;\alpha}+c_{i;E})-\mathcal J^{\rm aux}_i,
\label{eq:Jsource}
\end{align}
where $\mathcal J^{\rm aux}_i$ is an auxiliary source, used only to extract the propagator. The purely electromagnetic contribution reads
\begin{equation}
\mathcal L[A]
=
-\frac{\nu e^2}{4\pi\hbar}\,
\epsilon^{ij}A_i\partial_0 A_j
-
\tau\frac{\nu\alpha e^2\hbar\omega_c}{16\pi m^2v^3}
\epsilon^{ij} \partial_i A_j
\partial_0 A_y\, .
\label{eq:LA_A}
\end{equation}
The first term is the Chern--Simons response, which gives the quantized Hall conductivity, while the second is a tilt-induced magnetoelectric coupling. This coupling generates polarization and magnetization currents
that vanish in the spatially homogeneous limit.
Integrating out the internal gauge fluctuation \(\delta a_i\) while treating the $c_{ij;E}$-term perturbatively gives the
electromagnetic effective action (see SM \cite{sm}), 
\begin{align}
S_{\rm eff}[A]
&=
-i\hbar
\ln
\int \mathcal D[\delta a]\,
\exp\!\left[\frac{i}{\hbar}S[\delta a,A]\right]
\nonumber\\
&=
\frac{1}{2}
\int \frac{d^3q}{(2\pi)^3}\,
\chi^{(1)}_{ij}(-q,q)\,
A_i(-q)A_j(q)
\nonumber\\
&
+
\frac{1}{3}
\int \frac{d^3q_1}{(2\pi)^3}
\frac{d^3q_2}{(2\pi)^3}\,
\chi^{(2)}_{ijk}(-q_1-q_2,q_1,q_2)
\nonumber\\
&\times
A_i(-q_1-q_2)A_j(q_1)A_k(q_2)
+\cdots .
\label{eq:Seff}
\end{align}
where repeated spatial indices are summed and the frequency-conserving delta
functions are implicit. The kernels \(\chi^{(1)}\) and \(\chi^{(2)}\) encode,
respectively, the linear and second-order electromagnetic response generated by the composite-boson gauge fluctuations. For the general functional-derivative formulation of nonlinear electromagnetic response and the associated gauge-invariance constraints, see Ref.~\cite{rostami_ap_2021}.  The physical current is
obtained by functional differentiation,
\mbox{$j_i(q)=(2\pi)^3{\delta S_{\rm eff}[A]}/{\delta A_i(-q)}$} with 3-momentum $q=(\mathbf{q},\omega)$. For a diagrammatic representation of the different terms in the action, we refer to SM \cite{sm}. 

\PRLsection{Results and discussion}
We express the linear and nonlinear conductivity tensors entering the nonlinear quantum Hall resistivity in terms of
\(X\equiv\omega/\omega_c\) and \(Y\equiv\hbar\omega_c/\Delta\), which compare the drive, cyclotron, and gap scales. 
In the large-gap, weak-tilt regime, \(Y\ll1\) and \(\alpha\ll1\), the leading dynamical conductivity of a tilted massive Dirac system is,  
\begin{equation}
\label{eq:sigma1}
\sigma_{ij}(\omega)\simeq
\nu\frac{e^2}{h}
\left[
\epsilon_{ji}
+\frac{\epsilon_{ji}X^2+i\delta_{ij}X}{1-X^2}
\left(1+\frac{\tau Y}{2}\right)
\right].
\end{equation}
where $i,j\in\{x,y\}$. Note that the longitudinal part $\sim \delta_{ij}$ is purely reactive since our model does not include any loss.
Higher-order tilt corrections break rotational symmetry, leading to
\(\sigma_{xx}\neq\sigma_{yy}\), although \(\sigma_{xy}=-\sigma_{yx}\) is retained, with
the corresponding tensor components given in the Supplemental Material \cite{sm}.
Here \(\tau=\pm1\) labels the valley. 
In the lowest-Landau-level regime relevant below, the valley-Zeeman coupling polarizes the ground state, as shown in the lower panel of Fig.~\ref{fig:sketch}; we therefore set $\tau=-1$ for the electron-doped case from now on. 
The pole in the first-order conductivity, appearing at $X=1$ i.e. at $\omega=\omega_c$, is the center-of-mass Kohn mode~\cite{Kohn1961}.
Alternatively, it can be read from the zero of the inverse Green's function in Eq.~\eqref{eq:Ginv}, which also gives the usual dispersion $\omega_q \approx \omega_c + c q^2 $\cite{sm}  where the dispersion coefficient $c=\frac{\nu K}{2h}-\frac{\hbar^2\omega_c}{8(mv)^2} $ is controlled primarily by the compressibility \(K\), but with subleading corrections generated by the \(\pi^4\) nonlinear term in the Lagrangian.

We next compute the conductivity at $2\omega$ using diagrammatic perturbation theory for the composite-boson action within a finite-gap extension of the single-band GLCS theory, controlled for $Y\ll1$.
In the large-gap, weak-tilt regime, including permutation symmetry, the dominant second-order contribution is~\cite{sm}
\begin{equation}
\hspace{-2mm}\sigma^{(2)}_{ijk}(2\omega)
\simeq
\nu \frac{e^3}{h}\,
\frac{\alpha\hbar v}{4\Delta^2}\,
\frac{1-Y/2}
{(1-X^2)^2(1-4X^2)}
\,
\mathcal F_{ijk}(X),
\label{eq:sigma2}
\end{equation}
where the tensor structure is encoded in the polynomial functions
\(\mathcal F_{ijk}(X)\), listed explicitly in SM~\cite{sm}. 
The denominator in Eq.~\eqref{eq:sigma2} encodes Kohn-mode resonances:
a simple two-photon pole at $X=1/2$ ($2\omega=\omega_c$) and a
stronger double pole at $X=1$ ($\omega=\omega_c$), arising from
repeated single-frequency propagators. 
The remaining second-order diagrams are parametrically smaller, scaling as
\(\alpha^3/\Delta^{3}\), \(\alpha^3/\Delta^{6}\), and \(\alpha^5/\Delta^{4}\).
Keeping in mind that the expansion in $1/\Delta$ imposes a 
cutoff in the loop integration momenta, we can ask how our tree-level results are modified
by loops. This is discussed in some detail in the SM \cite{sm}, where we show that  
the nonlinear transport coefficients are dominated by the tree-level results.

\begin{figure}
    \centering
    \includegraphics[width=1\linewidth]{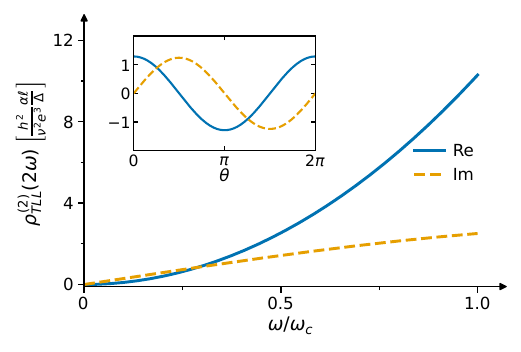}
\caption{
Nonlinear Hall resistivity \(\rho^{(2)}_{TLL}(2\omega)\) of a fractional quantum Hall liquid in the tilted Dirac model, in units of \(h^{2} \alpha\ell /(\nu^{2}e^{3}\Delta)\), with \(\ell=\hbar v/(8\Delta)\). The main panel shows the frequency dependence at \(\theta=\pi/4\). The inset shows the angular dependence at \(\omega/\omega_c=0.3\), using the same resistivity units. Solid blue and dashed orange curves denote the real and imaginary parts, respectively. Throughout, \(\hbar\omega_c/\Delta=0.1\) and \(\tau=-1\).
}
\label{fig:second_harmonic_hall_resistivity}
\end{figure}

Using $\rho_{ij}=(\sigma^{-1})_{ij}$ in Eq.~\eqref{eq:rho_TLL} and
rotating Eqs.~\eqref{eq:sigma1} and \eqref{eq:sigma2} from the $(x,y)$
frame to the longitudinal--transverse $(L,T)$ frame by an angle $\theta$
yields the linear and nonlinear resistivities. The full finite-frequency
expression for the second-order resistivity is given in the SM~\cite{sm}.
Figure~\ref{fig:second_harmonic_hall_resistivity} shows the frequency
dependence of the nonlinear Hall resistivity. Its real and imaginary
parts exhibit the leading low-frequency behaviors
$\operatorname{Re}\rho^{(2)}_{TLL}\propto\omega^2$ and
$\operatorname{Im}\rho^{(2)}_{TLL}\propto\omega$, respectively.
The response therefore vanishes in the dc limit. Notably, the
Kohn-mode resonances at $2\omega=\omega_c$ and $\omega=\omega_c$
present in the second-order conductivity cancel in the nonlinear
resistivity, leaving $\rho^{(2)}_{TLL}$ analytic throughout the
frequency range shown in
Fig.~\ref{fig:second_harmonic_hall_resistivity}.
Additional $Y$-dependent poles formally arise at much higher
frequencies from the linear-resistivity factors in
Eq.~\eqref{eq:rho_TLL}, but lie outside the low-frequency,
single-band regime considered here.
The inset shows the angular dependence, revealing a pronounced
anisotropy of the second-harmonic Hall response relative to the
crystalline tilt axis.
For $X=\omega/\omega_c\ll1$, $Y=\hbar\omega_c/\Delta\ll1$,
and $\alpha\ll1$, the low-frequency expansion of the second-order Hall resistivity reduces to Eq.~\eqref{eq:rho_qh_LT}, with 
\begin{align}
\begin{bmatrix}
  f(\theta)\\[3pt] g(\theta)  
\end{bmatrix}
\simeq \frac{\alpha\hbar v}{4\Delta^2}
\begin{bmatrix}
 \left(6-11 \frac{\hbar\omega_c}{\Delta}\right)\cos(\theta)
 \\[3pt]
 \left(2-\frac{\hbar\omega_c}{\Delta}\right)\sin(\theta)
\end{bmatrix} 
\label{eq:fg_result}
\end{align}
To examine the corresponding plateau structure, we consider the QH regime,
$|\rho_{LL}|,|\rho_{TT}|\ll|\rho_H|$, for which Eq.~\eqref{eq:rho_TLL} takes the approximate form
$
    \rho^{(2)}_{TLL}(2\omega)
    \simeq
    -\rho_H(2\omega)\,
    \sigma^{(2)}_{LTT}(2\omega)\,
    \rho_H^2(\omega)
    \propto \nu^{-2}
$,
where $\rho_H\simeq h/(\nu e^2)$ and
$\sigma^{(2)}\propto\nu$. This holds as long as
$2\omega\ll\omega_c$ and $2\hbar\omega\ll\Delta$. Strictly speaking, our impurity-free perturbative treatment about the mean-field state applies directly only at the center of the QH plateau, where $2\pi\hbar\bar{\rho}=\nu eB$. Away from the center, we assume that the density mismatch is accommodated by localized quasiparticles, leaving the bulk QH fluid unchanged. Under this assumption, $\rho^{(2)}_{TLL}(2\omega)$ remains pinned to its plateau-center value
over a finite density interval. Determining the width of this nonlinear plateau requires a treatment of disorder and localization beyond the
present approach. 

More broadly, our results complement recent studies of third-order
nonlinear response in graphene~\cite{He2024,Isobe2024} and suggest
extensions to higher-order nonlinearities, multicomponent Landau levels,
and moir\'e Chern systems~\cite{GirvinMacDonald1996,Cai2023}.
Finally, finite-frequency nonlinear Hall measurements may provide a
bulk probe of fractional quantum Hall fluids beyond dc transport.
 
\PRLsection{Acknowledgments}
This work was supported by \mbox{EPSRC} under Grant No.~UKRI122 and by the Royal Society under Grant Nos.~IES\textbackslash R2\textbackslash242309 and RGS\textbackslash R2\textbackslash252194. T.H.H. acknowledges support from the David Parkin Visiting Professorship Programme at the University of Bath and thanks the Department of Physics for its hospitality.

\bibliography{refs}   
\end{document}


\title{Supplemental Material for
``Second-Harmonic Bulk Response of Fractional Quantum Hall Fluids''}

\author{J.~Luke Pimlott}
\affiliation{
Department of Physics, University of Bath,
Claverton Down, Bath BA2 7AY, United Kingdom
}

\author{Pablo Reiser}
\affiliation{
Department of Physics, University of Bath,
Claverton Down, Bath BA2 7AY, United Kingdom
}

\author{T.~Hans Hansson}
\affiliation{
Fysikum, Stockholm University,
106 91 Stockholm, Sweden
}

\author{Habib Rostami}
\affiliation{
Department of Physics, University of Bath,
Claverton Down, Bath BA2 7AY, United Kingdom
}

\date{\today}

\maketitle

\vspace{-1cm}

\tableofcontents

\vspace{0.5cm}

In this Supplemental Material, we derive the effective low-energy Lagrangian used in the main text, construct the electromagnetic effective action by integrating out the Chern–Simons gauge fluctuations, and evaluate the linear and second-order conductivities. We also identify the diagrammatic origin of the dominant nonlinear response and show that loop corrections are subleading.

\section{Ginzburg-Landau Lagrangian Derivation for the Tilted Dirac System}

Here, we derive the low-energy Lagrangian density for the tilted
Dirac system in external electric and magnetic fields. After
bosonisation and inclusion of the Chern--Simons term~\cite{ZHK,ZHANG1992}, this forms the
effective theory used to describe the odd-denominator fractional
quantum Hall states considered in this work.

\subsection{Expansion of the full Lagrangian in the large gap regime}
We start from Eq.~(8) in the main text, substituting in the parameters specified just after in Eq.~(10). This describes the one-band projected Lagrangian (see also Son–Yamamoto \cite{Son2013} as a technical reference for the projection procedure), given by
\begin{equation}
    \mathcal{L}_{\text{eff}}=\psi^\dagger \left\{i\hbar\partial_t  + eA_0 - \alpha v\tau \pi_x - v^2\pi_{-}\big\{i\hbar\partial_t + eA_0 + 2\Delta - \alpha v\tau\pi_x\right\}^{-1}\pi_{+}\big\}\psi,
\end{equation}
where $v$ is the Dirac velocity, $\psi$ and $\psi^\dagger$ are the fermionic field operators, $2\Delta$ is the band gap and $\pi_i=p_i+eA_i$ is the $i$-component of the kinematic  momentum, with canonical momentum $p_i$, fundamental charge $e>0$ and electromagnetic vector potential $A_i$. Additionally, $\pi_{\pm}=\tau\pi_x \pm i\pi_y$, with valley index $\tau$. For the time being, we set $\hbar=v=1$ and drop the scalar potential term, $eA_0$, reintroducing it later, noting that it comes with the $i\partial_t$ term.

Expanding the resolvent in the final term for a large $\Delta$ as specified by Eq.~(9) in the main text, our Lagrangian up to second order in this expansion becomes
\begin{align}
    \mathcal{L}_{\text{eff}}=&\psi^\dagger \bigg\{i\partial_t  - \alpha\tau\pi_x - \frac{1}{2\Delta} \pi_{-}\pi_{+} + \frac{1}{(2\Delta)^2} \pi_{-}(i\partial_t - \alpha\tau\pi_x)\pi_{+} - \frac{1}{(2\Delta)^3} \pi_{-}(i\partial_t - \alpha\tau\pi_x)^2\pi_{+}\bigg\}\psi.
\end{align}
Expanding the parentheses, discarding the $\partial_t^2$ term and grouping the terms containing time derivatives, the Lagrangian is
\begin{align}
    \mathcal{L}_{\text{eff}}=&\psi^\dagger \bigg\{i\partial_t + \frac{1}{4\Delta^2} \pi_{-} i\partial_t \pi_{+} +\frac{\alpha\tau}{8\Delta^3} \pi_{-} i\partial_t \pi_x \pi_{+} + \frac{\alpha\tau}{8\Delta^3} \pi_{-} \pi_x i\partial_t \pi_{+} \nonumber \\
    &- \alpha\tau\pi_x - \frac{1}{2\Delta} \pi_{-}\pi_{+} - \frac{\alpha\tau}{4\Delta^2} \pi_{-} \pi_x \pi_{+} - \frac{\alpha^2}{8\Delta^3} \pi_{-} \pi_x \pi_x \pi_{+} \bigg\}\psi.
\end{align}
Using the identities
\begin{equation}
    \pi_{-} \partial_t \pi_+ = \frac{1}{2} (\partial_t \pi_- \pi_+ + \pi_- \pi_+ \partial_t + [\pi_- , \partial_t] \pi_+ + \pi_- [\partial_t ,\pi_+])
\end{equation}
in the second term, $\pi_-\partial_t=[\pi_-,\partial_t]+ \partial_t\pi_-$ in the third and $\partial_t \pi_+=[\partial_t,\pi_+]+ \pi_+ \partial_t$ in the fourth term of the Lagrangian, and note that the commutator terms yield electric fields: $[\partial_t ,\pi_i] = -[\pi_i, \partial_t] = -eE_i$, we obtain
\begin{align}
    \mathcal{L}_{\text{eff}}=&\psi^\dagger \bigg\{i\partial_t + \frac{1}{8\Delta^2} i\partial_t \pi_- \pi_+ + \frac{1}{8\Delta^2} \pi_- \pi_+ i\partial_t +\frac{\alpha\tau}{8\Delta^3} i\partial_t \pi_{-} \pi_x \pi_{+} + \frac{\alpha\tau}{8\Delta^3} \pi_{-} \pi_x \pi_{+} i\partial_t \nonumber \\
    &- \alpha\tau\pi_x - \frac{1}{2\Delta} \pi_{-}\pi_{+} - \frac{\alpha\tau}{4\Delta^2} \pi_{-} \pi_x \pi_{+} - \frac{\alpha^2}{8\Delta^3} \pi_{-} \pi_x \pi_x \pi_{+}\nonumber \\
    &+ \frac{ie}{8\Delta^2} E_- \pi_+ - \frac{ie}{8\Delta^2} \pi_- E_+ + \frac{ie\alpha\tau}{8\Delta^3} E_- \pi_x \pi_{+} - \frac{ie\alpha\tau}{8\Delta^3} \pi_{-} \pi_x E_+ \bigg\}\psi.
\end{align}
Up to order $\left(\frac{\pi}{\Delta}\right)^3$, the same order our resolvent was expanded to, the first line of the Lagrangian containing $\partial_t$ can be written in a simplified way:
\begin{align}
    &\psi^\dagger  \left(1 + \frac{1}{8\Delta^2} \pi_- \pi_+ + \frac{\alpha\tau}{8\Delta^3} \pi_{-} \pi_x \pi_{+} \right) i\partial_t \left(1 + \frac{1}{8\Delta^2} \pi_- \pi_+ + \frac{\alpha\tau}{8\Delta^3} \pi_{-} \pi_x \pi_{+} \right) \psi.
\end{align}
This allows us to renormalise our field operators as
\begin{align}
    \psi^{(\dagger)}\to \left(1 + \frac{1}{8\Delta^2} \pi_- \pi_+ + \frac{\alpha\tau}{8\Delta^3} \pi_{-} \pi_x \pi_{+} \right) \psi^{(\dagger)},
\end{align}
with the inverse of the renormalisation operator being
\begin{align}
    \left(1 + \frac{1}{8\Delta^2} \pi_- \pi_+ + \frac{\alpha\tau}{8\Delta^3} \pi_{-} \pi_x \pi_{+} \right)^{-1} =& 1 - \frac{1}{8\Delta^2} \pi_- \pi_+ - \frac{\alpha\tau}{8\Delta^3} \pi_{-} \pi_x \pi_{+} + \order{\pi^4}.
\end{align}
To order $\pi^4$, then, the Lagrangian takes the complicated form
\begin{align}
    \mathcal{L}_{\text{eff}}=&\psi^\dagger \bigg\{ i\partial_t + 
    \left(1 - \frac{1}{8\Delta^2} \pi_- \pi_+ - \frac{\alpha\tau}{8\Delta^3} \pi_{-} \pi_x \pi_{+} \right) \bigg(-\alpha\tau\pi_x - \frac{1}{2\Delta} \pi_{-}\pi_{+} - \frac{\alpha\tau}{4\Delta^2} \pi_{-} \pi_x \pi_{+} - \frac{\alpha^2}{8\Delta^3} \pi_{-} \pi_x \pi_x \pi_{+} \nonumber \\
    &+ \frac{ie}{8\Delta^2} E_- \pi_+ - \frac{ie}{8\Delta^2} \pi_- E_+ + \frac{ie\alpha\tau}{8\Delta^3} E_- \pi_x \pi_{+} - \frac{ie\alpha\tau}{8\Delta^3} \pi_{-} \pi_x E_+ \bigg) \left(1 - \frac{1}{8\Delta^2} \pi_- \pi_+ - \frac{\alpha\tau}{8\Delta^3} \pi_{-} \pi_x \pi_{+} \right)\bigg\}\psi,
\end{align}
which can be expanded in orders of $\pi$: 
\begin{equation}
    \mathcal{L}_{\text{eff}} = \mathcal{L}^{(2)}_{\text{eff}} + \mathcal{L}^{(3)}_{\text{eff}}+ \mathcal{L}^{(4)}_{\text{eff}} +{\cal O}(\pi^5). 
\end{equation}

\subsection{Up to second order in kinetic momentum: {\unboldmath$\mathcal{L}^{(2)}_{\rm eff}$}}

Up to order $\pi^2$, and reintroducing $\hbar$, $v$ and the $eA_0$ term, the Lagrangian is
\begin{align}
    \mathcal{L}^{(2)}_{\text{eff}}=&\psi^\dagger \bigg\{
     i\hbar\partial_t + eA_0- \alpha v\tau\pi_x - \frac{1}{2m} \pi_{-}\pi_{+} + \frac{i\hbar ev^2}{8\Delta^2} E_- \pi_+ - \frac{i\hbar e v^2}{8\Delta^2} \pi_- E_+ \bigg\}\psi,
\end{align}
using that $\Delta=mv^2$, with Dirac mass $m$.
We can also write this in terms of more familiar parameters, with identities $\pi_+ \pi_- = \pi_x^2 + \pi_y^2 + \hbar e\tau B$ and $E_- \pi_+ - \pi_- E_+ = 2i\tau (\mathbf{E}\times\mathbf{\pi})\cdot\hat{\mathbf{z}}$ (for spatially-homogeneous $\mathbf{E}$) allowing us to write
\begin{align}
    \mathcal{L}^{(2)}_{\text{eff}}=&\psi^\dagger \bigg\{
     i\hbar\partial_t + eA_0 + e\tilde{A}_0 - \frac{(-i\hbar\boldsymbol{\nabla}+e\mathbf{A}+e\tilde{\mathbf{A}})^2}{2m} - \boldsymbol{\mathcal{M}} \cdot \mathbf{B} - e\mathbf{E}\cdot\boldsymbol{\mathcal{A}(\pi)} \bigg\}\psi.
\end{align}
Here, the effective scalar and vector potentials $\tilde{A}_0=\frac{\alpha^2\Delta}{2e}$ and $\tilde{\mathbf{A}}=\frac{\alpha\Delta\tau}{ev}\hat{\mathbf{x}}$ describe the shift in the Dirac points due to the tilt, the magnetisation $\boldsymbol{\mathcal{M}} = \frac{\hbar e\tau}{2m} \hat{\mathbf{z}}$ couples the magnetic field and is responsible for the valley Zeeman effect, and the Berry connection $\boldsymbol{\mathcal{A}(\pi)} = \frac{\hbar v^2\tau}{4\Delta^2}(\pi_y,-\pi_x)$ couples the electric field and is responsible for the anomalous velocity. 
This form is useful for identifying the physical significance of each
term, although it is less convenient for the calculations below. Instead, we write this contribution in the main text as
\begin{align}
    \mathcal{L}^{(2)}_{\text{eff}}=&\psi^\dagger \bigg\{
     i\hbar\partial_t + eA_0- \alpha \tau v\pi_x - \frac{\abs{\boldsymbol{\pi}}^2}{2m} - \tau\frac{\hbar \omega_c}{2}  - \tau\frac{e\hbar}{4m^2v^2} \mathbf{E} \cdot (\boldsymbol{\pi} \times \hat{\mathbf{z}}) \bigg\}\psi,
\end{align}
where $\omega_c=\frac{eB}{m}$ is the cyclotron frequency and $\abs{\boldsymbol{\pi}}^2=\pi_x^2+\pi_y^2$.

\subsection{Terms third order in kinetic momentum: {\unboldmath$\mathcal{L}^{(3)}_{\mathrm{eff}}$}}

Next, the $\pi^3$-order contribution to the Lagrangian is
\begin{align}
    \mathcal{L}^{(3)}_{\text{eff}}=&\frac{\alpha v^3\tau}{8\Delta^2} \psi^\dagger\bigg\{\pi_- \pi_+ \pi_x + \pi_x \pi_- \pi_+  - 2 \pi_{-} \pi_x \pi_{+} + \frac{i\hbar e}{\Delta} (E_- \pi_x \pi_{+} - \pi_{-} \pi_x E_+) \bigg\}\psi.
\end{align}
After some algebra, including the use of $[\pi_y,\pi_x]=i\hbar eB$, we can write this as
\begin{align}
    \mathcal{L}^{(3)}_{\text{eff}}=& \psi^\dagger\bigg\{- \frac{\hbar \omega_c}{4\Delta}\alpha v\pi_x 
    - \frac{e\hbar\alpha}{4(mv)^3} \left(E_x \frac{\{\pi_y, \pi_x\}}{2} - E_y \pi_x^2 \right) + \tau\frac{\alpha (e\hbar)^2}{8(mv)^3}E_yB \bigg\}\psi,
\end{align}
which is how we write this in the main text. The first term can be seen as a higher-order magnetic correction to the tilt ($\alpha$-dependent) term of $\mathcal{L}^{(2)}_{\text{eff}}$, while the third is a valley-dependent direct magnetoelectric coupling term. The second term, however, is the source of the leading-order nonlinear conductivity which we are interested in, which we later label as the $c_{ij;E}$ term.

\subsection{Leading fourth order terms in kinetic momentum: {\unboldmath$\mathcal{L}^{(4)}_{\mathrm{eff}}$}}
Finally, we look at the $\pi^4$-order contribution to the Lagrangian, taking only the leading-order part varying as $\Delta^{-3}$. We retain this contribution to capture the leading nonparabolic correction to the projected band dispersion. The relevant part of the Lagrangian is
\begin{align}
    \mathcal{L}^{(4)}_{\text{eff,L}}=&\frac{v^4}{8\Delta^3}\psi^\dagger\bigg\{\pi_{-}\pi_{+} \pi_{-}\pi_{+} + \alpha^2 \Big(\pi_{-} \pi_x \pi_{+} \pi_x + \pi_x \pi_{-} \pi_x \pi_{+} - \pi_{-} \pi_x \pi_x \pi_{+}\Big) \bigg\}\psi.
\end{align}
Expanding $\pi_\pm$ into their Cartesian components and performing similar algebra to the previous terms, one ends up with
\begin{align}
    \mathcal{L}^{(4)}_{\text{eff,L}}=&\psi^\dagger\bigg\{ \frac{\abs{\boldsymbol{\pi}}^4}{8m^3v^2} + \tau\frac{\hbar\omega_c}{2\Delta}\frac{\abs{\boldsymbol{\pi}}^2}{2m} + \frac{\hbar^2 \omega_c^2}{8\Delta} + \alpha^2 \Big(\frac{\pi_x^4 + \pi_y\pi_x^2\pi_y}{8m^3v^2} + \tau\frac{\hbar\omega_c}{4\Delta} \frac{\pi_x^2}{2m} - \frac{\hbar^2 \omega_c^2}{8\Delta} \Big) \bigg\}\psi.
\end{align}
The higher-order terms include magnetic and tilt-dependent corrections to the parabolic kinetic energy, $\abs{\boldsymbol{\pi}}^2/2m$. The tilt-dependent corrections distinguish $\pi_x$, consistent with the tilt direction. We also obtain Zeeman corrections of second order in $\omega_c$ that are valley-independent, unlike the first-order valley-Zeeman term. In the regime $\alpha\ll 1$ and $\hbar\omega_c\ll\Delta$, we neglect these corrections except for the first term, proportional to $\abs{\boldsymbol{\pi}}^4$, which is the leading relativistic-like  correction to the parabolic kinetic energy.
Combining the retained contributions gives the single-particle Lagrangian density quoted in Eq.~(11) of the main text, valid to the order required for our analysis. These contributions comprise: (i) the parabolic kinetic energy and valley-Zeeman coupling; (ii) the anomalous-velocity, or Berry-connection, coupling to the electric field; (iii) the leading tilt-induced cubic term, which generates the nonlinear Hall response; and (iv) the leading quartic kinetic correction, which captures
the first nonparabolic correction to the projected band dispersion. All remaining terms are neglected because they are higher order in $\alpha$ or $\hbar\omega_c/\Delta$, or do not contribute to the leading nonlinear conductivity.

\section{Electromagnetic effective action after integrating out gauge fluctuations}

In this section, we perform the Gaussian integration in the Gaussian partition function describing our bosonised tilted Dirac system over the Chern-Simons statistical field fluctuations, $\delta\mathbf{a}$. This leaves us with a partition function that is dependent on the external field, which with the leading-order nonlinear correction to the action treated perturbatively gives us the effective action encompassing the electromagnetic response of the system, including nonlinear orders. We assume Einstein summation convention over repeated indices throughout.

\subsection{Perturbative expansion of the effective action}

We start from the action given by Eq.~(18) in the main text, with the
$c_{ij;E}$ term being treated perturbatively below. The remaining action is
Gaussian in the statistical gauge-field fluctuation $\delta a$. Integrating over $\delta a$ gives the Gaussian generating functional \cite{AltlandSimons} 
\begin{align}
Z_G[j,A]
\equiv
\int D[\delta a]\,
\exp\left[
\frac{i}{\hbar}S[\delta a,A,j]
\right]
 =
\mathcal {N}_G
\exp\left\{
\frac{i}{\hbar}
\int d^3x\,
\left[
\frac{1}{2}\mathcal{J}_i G^{ij}\mathcal{J}_j
-\mathcal{L}[A]
\right]
\right\},
\label{eq.SM_ZG}
\end{align}
where the field-independent Gaussian normalization \(\mathcal N_G\) has been suppressed and $\mathcal{J}_i$ contains the
auxiliary source $j_i$, denoted
$\mathcal{J}^{\rm aux}_i$ in the main text, and $G^{ij}$ is the inverse of the Gaussian
kernel, $[G^{-1}]_{ij}$. 
Neglecting the fourth-order spatial-derivative term in the Gaussian kernel of Eq.~(19) and inverting the resulting kernel, we obtain the Green's function
\begin{equation}
G^{ij}
=\frac{h}{\nu}
\frac{
\omega_c\delta^{ij}
+\epsilon^{ij} \partial_0
-\frac{\nu K}{h}\partial_i\partial_j
+\frac{\hbar^2\omega_c}{4m^2v^2}
\epsilon^{ki}\epsilon^{lj}\partial_k\partial_l
}{
\omega_c^2+\partial_0^2
-\left(
\frac{\nu K\omega_c}{h}
-\frac{\hbar^2\omega_c^2}{4m^2v^2}
\right)\nabla^2
-\frac{\nu K\hbar\omega_c}{8\pi  m^2v^2}\nabla^4}. 
\label{eq.SM_Gij}
\end{equation}
We now restore the interaction term omitted from the Gaussian integration,
\begin{equation}
S_{\mathrm{int}}[\delta a]
=
\int d^3x\,\mathcal{L}_{\mathrm{int}}[\delta a].
\label{eq.SM_Sint}
\end{equation}
Adding this term to the Gaussian action gives
\begin{align}
Z[j,A]
&=
\int D[\delta a]\,
\exp\left[
\frac{i}{\hbar}
\left(
S[\delta a,A,j]
+
S_{\mathrm{int}}[\delta a]
\right)
\right]
=
\exp\left[
\frac{i}{\hbar}\widehat S_{\mathrm{int}}
\right]
Z_G[j,A].
\label{eq.SM_Z_interacting}
\end{align}
with the corresponding functional differential operator 
\begin{equation}
\widehat S_{\mathrm{int}}
\equiv
\int d^3x\,
\mathcal{L}_{\mathrm{int}}
\left[
\frac{\hbar}{i}\frac{\delta}{\delta j(x)}
\right]. 
\label{eq.SM_Sint_operator}
\end{equation}
The physical generating functional is obtained by setting $j=0$ after
the functional derivatives have been evaluated. Defining the effective
action through $e^{\frac{i}{\hbar}S_{\mathrm{eff}}[A]}
=
\left.Z[j,A]\right|_{j=0}$ and expanding perturbatively in $S_{\mathrm{int}}$, the first-order result is
\begin{equation}
S_{\mathrm{eff}}[A]
=
-i\hbar\ln Z_G[0,A]
+
Z^{-1}_G[0,A]\, 
\widehat S_{\mathrm{int}}\, Z_G[j,A]
\Big|_{j=0}\,
+
\order{S_{\mathrm{int}}^2}.
\label{eq.SM_Seff1_general}
\end{equation}

This contains the linear fractional quantum Hall response and its
finite-frequency and tilt-dependent tree-level corrections, as well as
the leading tree-level contribution to the second-order conductivity.
For completeness, the second-order contribution is
\begin{align}
S_{\mathrm{eff}}^{(2)}[A]
=
\frac{i}{2\hbar}
\left[
Z^{-1}_G[0,A]
\widehat S_{\mathrm{int}}^2 Z_G[j,A]
\Big|_{j=0}
-
\left(
Z^{-1}_G[0,A]\widehat S_{\mathrm{int}} Z_G[j,A]
\Big|_{j=0}
\right)^2
\right].
\label{eq.SM_Seff2ndOrdPert}
\end{align}
The first term generates the second-order perturbative contributions,
while the second subtracts the disconnected part. This produces
loop corrections to the linear and second-order conductivities, together
with the leading tree-level contribution to the third-order conductivity.
We use it below to identify the loop contributions and their
order in the perturbative expansion.
In our case, the interaction part of the action arises from the \(c_{ij,E}\) term:
\begin{align}
    \mathcal{L}_{\text{int}}[\delta a] =& - \frac{\nu e\alpha\omega_c}{8\pi m^2 v^3} \left(E_x \delta a_y \delta a_x - E_y \delta a_x \delta a_x \right),
\end{align}
where we make the substitution \(\pi \to \delta a\) and omit derivative terms. These are momentum-dependent contributions that either vanish in the quantities considered below or are parametrically negligible in the regime of interest.

\subsection{Fourier-space expression for the effective action}
From here, we perform all our analysis in frequency--momentum space rather than real spacetime, meaning we use the Fourier-transform expressions for $Z_G$ and the integrated $\mathcal{L}_{\text{int}}$:
\begin{align}
    Z_G[j,A] =& \exp[\frac{i}{\hbar} \int \frac{d^3q}{(2\pi)^3} \, \left[\frac{1}{2} \mathcal{J}_i[j,A,-q] G^{ij}(q) \mathcal{J}_j[j,A,q] - \mathcal{L}[A,q]\right]]
\end{align}
and
\begin{align}
    \int d^3x \mathcal{L}_{\text{int}}[\delta a] =& - \frac{\nu e\alpha\omega_c \epsilon^{\alpha\beta}}{8\pi m^2 v^3}  \int \frac{d^3 q_a  d^3 q_b}{(2\pi)^6}\, E_\alpha(-q_a-q_b) \delta a_\beta(q_a) \delta a_x(q_b), \label{eq.SM_intx_L_int}
\end{align}
where we replace \(\delta a_i(q)\) by
\begin{equation}
   \delta a_i(q)\to -i\hbar(2\pi)^3\frac{\delta}{\delta j_i(-q)},
\end{equation}
the Fourier-space representation of \(-i\hbar\,\delta/\delta j_i(x)\).
The Fourier-space expressions of the propagator follow
\begin{align}
    G^{ij}(q) = \frac{h}{\nu} \frac{ \omega_c\delta^{ij}+i\omega \epsilon^{ij} + \frac{\nu K}{h}q_i q_j - \frac{\hbar^2 \omega_c}{4(mv)^2}\epsilon^{ki}\epsilon^{lj} q_k q_l}{\left(\omega_c^2 -\omega^2 \right) + \left(\frac{\nu K \omega_c}{h} - \frac{\hbar^2 \omega_c^2}{4m^2 v^2}\right)|\mathbf{q}|^2 - \frac{\nu K\hbar\omega_c}{8\pi m^2v^2} |\mathbf{q}|^4 }, \label{eq.SM_Gij_q}
\end{align}
where $q_i$ is a component of the spatial momentum, while $\omega$ is the temporal component of $q$; the total source term is
\begin{equation}\label{eq.SM_Ji}
   \mathcal J_i[j,A,q]  =\mathcal{J}^{(0)}_i[A,q] - j_i(q),
\end{equation}
with
\begin{align}
\mathcal{J}^{(0)}_i[A,q]=
\tau_B \frac{\nu\alpha mv\omega_c}{2\pi\hbar} (2\pi)^3\delta(q)\delta_{ix} - \frac{\tau_B}{\tau}\frac{i\nu e \omega}{2\pi\hbar} \epsilon^{ji} A_j(q),
\end{align}
in which we define
\begin{equation}
    \tau_B=\tau+\frac{\hbar\omega_c}{4\Delta};
\end{equation} 
and the purely electromagnetic contribution
\begin{align}
    \mathcal{L}[A,q] = -A_i(-q) \frac{i\nu e^2 \omega}{4\pi\hbar} \varepsilon^{ij} A_j(q) -
\tau\frac{i\nu\alpha e\hbar\omega_c^2\omega}{16\pi mv^3} A_y(q). \label{eq.SM_L}
\end{align}
Substituting these expressions into Eq.~\eqref{eq.SM_Seff1_general}, we obtain the zeroth-order contribution to the effective action,
\begin{align}
    S^{(0)}_{\text{eff}}[A]=- i\hbar\ln Z_G[0,A] =  \int \frac{d^3q}{(2\pi)^3} \, \left[\frac{1}{2} \mathcal{J}^{(0)}_i[A,-q] G^{ij}(q) \mathcal{J}^{(0)}_j[A,q] - \mathcal{L}[A,q]\right]. \label{eq.SM_Seff0}
\end{align}
At first order in the perturbative expansion, the effective-action
contribution is
\begin{align}
S_{\mathrm{eff}}^{(1)}[A]
=
\frac{
\left.
\widehat S_{\mathrm{int}}\,Z_G[j,A]
\right|_{j\to0}
}{
Z_G[0,A]
}
 =
\frac{\nu e\alpha\hbar^2\omega_c}{8\pi m^2v^3}\,
\frac{1}{Z_G[0,A]}\epsilon^{\mu\beta}
\displaystyle
\int d^3q_a\,d^3q_b\,
E_\mu(-q_a-q_b)\,
\left.
\frac{\delta}{\delta j_\beta(-q_a)}
\frac{\delta}{\delta j_x(-q_b)}
Z_G[j,A]
\right|_{j\to0}
.
\label{eq.SM_Seff1}
\end{align}
Using
\begin{equation}
\frac{\delta j_i(q)}{\delta j_j(q')}
=
\delta_{ij}\,\delta(q-q'),
\qquad
G^{ij}(q)=G^{ji}(-q). 
\label{eq.SM_functional_identities}
\end{equation}
The second functional derivative is
\begin{align}
\frac{\delta}{\delta j_\beta(-q_a)}
\frac{\delta}{\delta j_x(-q_b)}
Z_G[j,A]
=
\Bigg[
\frac{i}{\hbar(2\pi)^3}
\delta(q_a+q_b)\,
G^{x\beta}(q_b)
-\frac{1}{\hbar^2(2\pi)^6}
G^{xi}(q_b)\,
\mathcal{J}_i[j,A,q_b]\,
G^{\beta j}(q_a)\,
\mathcal{J}_j[j,A,q_a]
\Bigg]
Z_G[j,A].
\label{eq.SM_ZG_second_derivative}
\end{align}
Substituting this result into Eq.~\eqref{eq.SM_Seff1} and setting $j=0$
gives
\begin{align}
S_{\mathrm{eff}}^{(1)}[A]
&=
\frac{i\nu e\alpha\hbar\omega_c}{8\pi m^2v^3}
\epsilon^{\mu\beta}
\int \frac{d^3q}{(2\pi)^3}\,
E_\mu(0)\,
G^{x\beta}(q)
\nonumber\\
&-
\frac{\nu e\alpha\omega_c}{8\pi m^2v^3}
\epsilon^{\mu\beta}
\int \frac{d^3q_a\,d^3q_b}{(2\pi)^6}\,
E_\mu(-q_a-q_b)
G^{xi}(q_b)\,
\mathcal{J}^{(0)}_i[A,q_b]\,
G^{\beta j}(q_a)\,
\mathcal{J}^{(0)}_j[A,q_a].
\label{eq.SM_Seff1_explicit}
\end{align}

Examining the dependence of \(\mathcal{J}^{(0)}_j[A,q]\) and \(\mathcal{L}[A,q]\) on the gauge field, we find that \(S^{(0)}_{\mathrm{eff}}[A]\) contains terms up to \(\mathcal{O}(A^2)\), whereas \(S^{(1)}_{\mathrm{eff}}[A]\) contains terms up to \(\mathcal{O}(A^3)\), with \(E\sim\mathcal{O}(A)\). Since the current is obtained by differentiating the effective action with respect to \(A\), an \(\mathcal{O}(A^n)\) term in the action generates an \(\mathcal{O}(A^{n-1})\) current response. Thus, the quadratic and cubic terms determine the linear and second-order responses, respectively.
Terms below \(\mathcal{O}(A^2)\) can be omitted: the \(\mathcal{O}(A^0)\) term gives no current, while the \(\mathcal{O}(A^1)\) term corresponds to an equilibrium current, which vanishes in the present system. Therefore, at this order in the perturbative expansion, we may write
\begin{align}
    S_{\text{eff}}[A] &= S_{\text{eff}}^{(0)}[A] + S_{\text{eff}}^{(1)}[A]  +\cdots \\
    &= \frac{1}{2}\int \frac{d^3q}{(2\pi)^3}\,\chi^{(1)}_{ij}(-q,q) A_i(-q)A_j(q) + \frac{1}{3}\int \frac{d^3q_a}{(2\pi)^3}\,  \int \frac{d^3q_b}{(2\pi)^3}\, \chi^{(2)}_{ijk}(-q_a-q_b, q_a,q_b) A_i(-q_a-q_b)A_j(q_a)A_k(q_b) +\cdots \nonumber,
\end{align}
with kernels $\chi^{(n)}$ describing the $n$th-order electromagnetic response function. For the specific case of $n$th-harmonic responses, one should set all momenta equal through Dirac delta functions in each $\chi^{(n)}$, in which case the effective action becomes the one written in Eq.~(22) of the main text.
The physical current is obtained by differentiating the effective action with respect to the external vector potential,
\begin{equation}
  J_i(q) = (2\pi)^3 \frac{\delta S_{\rm eff}[A]}{\delta A_i(-q)} .
\end{equation}
We therefore define the response functions through the expansion
\begin{align}
    J_i(q) = \chi^{(1)}_{ij}(-q,q) A_j(q) + \int \frac{d^3q_1}{(2\pi)^3}\, \chi^{(2)}_{ijk}(-q, q_1, q-q_1) A_j(q_1)A_k(q-q_1) +\cdots \label{eq.SM_J_expansion_A}
\end{align}
where the linear and second-order response functions are defined by \cite{rostami_ap_2021}
\begin{align}
\chi^{(1)}_{ij}(-q,q)
&=
\left.
(2\pi)^3
\frac{\delta^2 S_{\rm eff}[A]}
{\delta A_i(-q)\,\delta A_j(q)}
\right|_{A=0},
\nonumber\\
\chi^{(2)}_{ijk}(-q_1-q_2,q_1,q_2)
&=
\left.
\frac{(2\pi)^6}{2}
\frac{\delta^3 S_{\rm eff}[A]}
{\delta A_i(-q_1-q_2)\,
 \delta A_j(q_1)\,
 \delta A_k(q_2)}
\right|_{A=0}.
\label{eq.SM_response_action_definitions}
\end{align}
The zero-order and first-order perturbation terms of $S_{\text{eff}}[A]$ describe the tree-level corrected first-order current and the leading-order tree-level second-order current, both of which we explicitly obtain below.

\subsection{Cyclotron resonance and the Kohn mode}
The pole of the composite-boson Green's function in Eq.~\eqref{eq.SM_Gij_q} signals the existence of a collective excitation, known as the Kohn mode \cite{Kohn1961}, whose leading-order dispersion is given by
\begin{align}\label{eq.SM_Kohn_mode}
    \omega^2_{\rm K}(q) =\omega_c^2  + \left(\frac{\nu K \omega_c}{h} - \frac{v^2\hbar^2 \omega_c^2}{4\Delta^2}\right) q^2  +{\cal O}(q^4),
\end{align}
which gives the finite-momentum dispersion of the Kohn mode. As we show below, both the linear and second-order dynamical conductivities exhibit cyclotron resonances associated with one- and two-photon excitation processes, respectively, at $\omega=\omega_c$. These resonances correspond to optical excitation of the collective Kohn mode rather than individual inter-Landau-level transitions. The corresponding poles in the optical conductivities therefore probe the $\mathbf{q}=0$ limit of the Kohn mode. Expanding its dispersion at small momentum yields 
\begin{equation}
\omega_{\rm K}(q) \approx
\omega_c
+
\left(
\frac{\nu K}{2h}
-
\frac{v^2\hbar^2\omega_c}{8\Delta^2}
\right) q^2,
\label{eq.SM_Kohn_small_q}
\end{equation}
as discussed in the main text. It is important to note that the curvature of the dispersion
remains positive within the validity regime of our projection,
characterised by $\hbar\omega_c/\Delta \ll 1$ and positive
inverse compressibility, $K>0$.

\section{Diagrammatic organisation and perturbative hierarchy}

\begin{figure}[ht]
    \centering
    \includegraphics[width=0.8\linewidth]{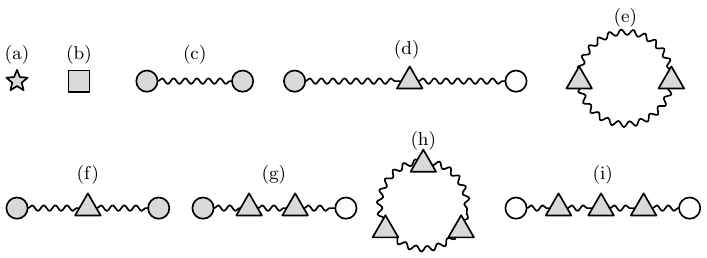}
    \caption{Diagrams contributing to the first-order current (top row) and second-order current (bottom row) obtained from the perturbative expansion. The star denotes the Chern--Simons vertex, responsible for the static fractional quantum Hall response, and the square denotes the magnetoelectric vertex, which generates a current proportional to the magnetic field, each coming from the respective term of $\mathcal{L}[A]$ (Eq.~\eqref{eq.SM_L}). Solid circles and open circles come from the effective source term $\mathcal{J}^{(0)}_i[A,q]$ (Eq.~\eqref{eq.SM_Ji}), and triangles come from the perturbing term of Eq.~\eqref{eq.SM_intx_L_int}. These latter three vertex functions are explicitly given by Eq.~\eqref{eq.VertexFuncs}. Wavy lines denote composite-boson propagators. (b) is exceptional in that it represents the current induced solely by the magnetic field and does not contribute to the linear conductivity. (c) and (f) give the leading tree-level linear and second-order responses. (d), (g), and (i) are subleading tree-level corrections, while (e) and (h) are loop corrections to the linear and second-order responses, respectively.
    }
    \label{fig:diagrams}
\end{figure}

We now recast the perturbative expansion of the effective action in
diagrammatic form. Figure~\ref{fig:diagrams} shows the contributions to
the linear and second-order currents in diagrams (a--e) and (f--i),
respectively. Diagrams (a) and (b) arise directly from $\mathcal{L}[A]$ and
contain no propagators, while the remaining diagrams represent tree-level
or loop contributions. In particular, (c) and (f) give the leading
tree-level linear and second-order responses, (d), (g), and (i) are
subleading tree-level corrections, and (e) and (h) are loop corrections.

The vertices scale differently with the tilt $\alpha$ and gap $2\Delta$
($m\propto\Delta$). The star and solid-circle vertices scale as
$\alpha^0\Delta^0$, the square as $\alpha/\Delta$, the open circle as
$\alpha\Delta$, and the triangle as $\alpha/\Delta^2$. We do not count
the implicit $\Delta$ dependence through $\omega_c=eB/m$.

Below, we evaluate the linear conductivity from diagrams (a), (c), and
(d), and show that the loop correction (e), although cutoff dependent, is subleading.
For the second-order conductivity, we evaluate the leading contribution
(f); the remaining contributions (g--i) are subleading.

The diagrams relate to the relevant expressions for their respective contribution through the Feynman rules, which are obtained through manual evaluation of the effective action to different orders. Each vertex relates to the element of a tensorial vertex function, with the indices and momenta at which they are evaluated also determined by the Feynman rules.

The star and square vertices appearing in the single-vertex diagrams (a) and (b) are respectively the two terms of $\mathcal{L}[A,q]$ of Eq.~\eqref{eq.SM_L}. The hollow circle, filled circle and triangle vertex are respectively determined by the vector, matrix and third-rank tensor vertex function elements
\begin{align}
    C_i(q) = \tau_B \frac{\nu\alpha mv\omega_c}{2\pi\hbar} (2\pi)^3\delta(q)\delta_{ix}; \hspace{0.7cm}
    F_{i;j}(q) = -\frac{\tau_B}{\tau}\frac{i\nu e}{2\pi\hbar} \omega \epsilon^{ji} \hspace{0.35cm} \text{and}\hspace{0.35cm}
    T_{ij;k}(q) = \frac{-i\nu e\hbar^2 \alpha\omega_c}{8\pi m^2 v^3}\omega(\delta^{il}\delta^{jx}+\delta^{jl}\delta^{ix})\epsilon^{kl}, \label{eq.VertexFuncs}
\end{align}
originating from the $A$-independent and $A$-dependent parts of $\mathcal{J}^{(0)}_i[A,q]$ (Eq.~\eqref{eq.SM_Ji}), and from $\int d^3x \mathcal{L}_{\text{int}}\left[\delta/\delta j\right]$ (see Eq.~\eqref{eq.SM_intx_L_int}), respectively.
The latter two come with an index after a semi-colon, which implies that in the action, they come with an external vector potential field with that index and momentum, such as $F_{i;j}(q) A_j(q)$. There is also a summation over repeated index $l$ in $T_{ij;k}(q)$.

Along with these rules for the vertices, how they are constructed with the propagators into the diagrams are subject to Feynman rules. One rule is momentum and index conservation and continuity at each vertex, such as $...G^{\lambda i}(q_2)T_{ij;k}(q_2-q_1)G^{j\beta}(q_1)...$ which, read right-to-left indicates incoming momentum $q_2-q_1$ and a shift from Cartesian index $j$ to index $i$ at the triangle vertex. Another rule is that we have as many independent space-time momenta, $q$ as propagators, all of which must be integrated over with a scaling of $(2\pi)^{-3}$. The single-vertex diagrams (a) and (b) also contain one integrated-over momentum despite containing no propagators. Additionally, diagrams come with an overall prefactor of
\begin{align}
    \left(\frac{i}{\hbar}\right)^{n_P+n_T-1}\times\frac{2^{n_P-3+\delta_{n_P,1}}}{n_T!}, \label{eq.PrefactorFeynRule}
\end{align}
where $n_P$ and $n_T$ are the number of propagators and triangle vertices, respectively. There is also an additional rule for diagrams with one filled and one hollow circle, stating that one must account for both permutations of how these two are ordered at the ends of the propagator chain.

\section{Linear conductivity}
Here, we evaluate the linear current obtained from the $\order{A^2}$ terms in \(S_{\mathrm{eff}}[A]\). The physical conductivity is extracted by differentiating the effective action with respect to the external vector potential and then expressing the result in terms of the electric field.

\subsection{From the zero-order perturbation}

We start with $S^{(0)}_{\text{eff}}[A]$, given by Eq.~\eqref{eq.SM_Seff0}, and represented diagrammatically by Figs. \ref{fig:diagrams}(a) and (c). Using the Feynman rules, this contribution is given by
\begin{align}
    S^{(0)}_{\text{eff}}[A^2] =  -\int \frac{d^3q}{(2\pi)^3} \,\mathcal{L}[A^2,q] + \frac{1}{2}\int \frac{d^3q}{(2\pi)^3} \,A_i(-q)F_{k;i}(-q) G^{kl}(q) F_{l;j}(q) A_j(q).
\end{align}
Using the vertex expressions for $\mathcal{L}[A,q]$ and $F_{l;j}(q)$ of Eqs.~\eqref{eq.SM_L} and \eqref{eq.VertexFuncs}, respectively, this evaluates to
\begin{align}
    S^{(0)}_{\text{eff}}[A^2] = \int \frac{d^3q}{(2\pi)^3} \, A_i(-q)\left[\frac{i\nu e^2 \omega}{4\pi\hbar} \varepsilon^{ij} + \frac{\tau_B^2\nu^2 e^2 \omega^2}{8\pi^2\hbar^2} \epsilon^{ik} \epsilon^{jl} G^{kl}(q) \right] A_j(q).
\end{align}
From the zero-order effective action, using ${\delta A_a(q)}/{\delta A_i(q')}=\delta^{ai}\delta(q-q')$, the first-order current is given by
\begin{align}
    J_a^{(1;0)}(q) = \left[\frac{i\nu e^2 \omega}{4\pi\hbar} \varepsilon^{aj} + \frac{\tau_B^2\nu^2 e^2 \omega^2}{8\pi^2\hbar^2} \epsilon^{ak} \epsilon^{jl} G^{kl}(q) \right] A_j(q) + A_i(q)\left[\frac{-i\nu e^2 \omega}{4\pi\hbar} \varepsilon^{ia} + \frac{\tau_B^2\nu^2 e^2 \omega^2}{8\pi^2\hbar^2} \epsilon^{ik} \epsilon^{al} G^{kl}(-q) \right],
\end{align}
which after some index manipulation results in the two permutations being equal, such that
\begin{align}
    J_a^{(1;0)}(q) = \left[\frac{i\nu e^2 \omega}{2\pi\hbar} \varepsilon^{aj} + \frac{\tau_B^2\nu^2 e^2 \omega^2}{4\pi^2\hbar^2} \epsilon^{ak} \epsilon^{jl} G^{kl}(q) \right] A_j(q),
\end{align}
where the contents of these square brackets  contribute to $\chi^{(1)}_{ij}(-q,q)$ at the leading order.
We consider a spatially homogeneous electric field,
$E(q)=-i\omega A(q)$, and hence restrict to zero spatial momentum,
$\mathbf{q}=0$. Accordingly,
\begin{equation}
    E(q)=(2\pi)^2 E(\omega)\delta(\mathbf{q}), 
\end{equation}
with an analogous expression for $J_a^{(1;0)}(q)$. Performing the
integration over $\mathbf{q}$ then gives
\begin{align}
    J_a^{(1;0)}(\omega) = \left[-\frac{\nu e^2}{2\pi\hbar} \varepsilon^{aj} + \frac{i\tau_B^2\nu^2 e^2 \omega}{4\pi^2\hbar^2} \epsilon^{ak} \epsilon^{jl} G^{kl}(\omega) \right] E_j(\omega),
\end{align}
with the propagator at zero spatial momentum taking on a much more simplified form:
\begin{align}
    G^{ij}(\omega) =&\frac{2\pi\hbar [\omega_c\delta^{ij}+i\omega \epsilon^{ij}]}{\nu\left(\omega_c^2 -\omega^2 \right)} 
    . \label{eq.SM_Gij_omega}
\end{align}

The conductivity $\sigma$ relates the induced current to the electric
field rather than to the vector potential. At linear order,
\begin{equation}
J_a(\omega)
=
\sigma^{(1)}_{ab}(\omega)E_b(\omega),
\label{eq.SM_linear_current}
\end{equation}
with
\begin{equation}
\sigma^{(1)}_{ab}(\omega)
=
\frac{\chi^{(1)}_{ab}(-\omega,\omega)}{-i\omega}.
\label{eq.SM_linear_conductivity_relation}
\end{equation}
Thus, at this order in the perturbative expansion, the longitudinal
conductivity components are
\begin{align}
    \sigma_{xx}^{(1;0)}(\omega) = \frac{\nu e^2}{h}\frac{i\tau_B^2 \omega_c \omega}{\omega_c^2 -\omega^2} \hspace{1cm} \text{and} \hspace{1cm} \sigma_{yy}^{(1;0)}(\omega) = \frac{\nu e^2}{h}\frac{i\tau_B^2 \omega_c\omega}{\omega_c^2 -\omega^2},
\end{align}
while the linear Hall conductivities are
\begin{align}
    \sigma_{xy}^{(1;0)}(\omega) =- \frac{\nu e^2}{h}\left(1 +\frac{\tau_B^2 \omega^2}{\omega_c^2 -\omega^2}\right) \hspace{1cm} \text{and} \hspace{1cm} \sigma_{yx}^{(1;0)}(\omega) = \frac{\nu e^2}{h} \left(1+\frac{\tau_B^2 \omega^2}{\omega_c^2 -\omega^2}\right).
\end{align}
These leading-order contributions to the linear conductivity are given in a unified and non-dimensionalised form in Eq.~(23) of the main text, where we also use $\tau_B^2\approx 1+\frac{\tau\hbar\omega_c}{2\Delta}$ for $\Delta\gg\hbar\omega_c$.

\subsection{From the first-order perturbation}

We now consider the tree-level correction to the linear conductivity
arising from the $\order{A^2}$ terms in the first-order contribution to
the effective action, $S^{(1)}_{\rm eff}[A]$, given in
Eq.~\eqref{eq.SM_Seff1}. This contribution is represented by
Fig.~\ref{fig:diagrams}(d) and involves both the $A$-dependent and
$A$-independent parts of $\mathcal{J}^{(0)}_i[A,q]$. From the Feynman rules, this contribution is
\begin{align}
     S_{\text{eff}}^{(1)}[A^2] 
    =& \frac{i^2}{2\hbar^2}\int \frac{d^3q_a}{(2\pi)^3} \int \frac{d^3q_b}{(2\pi)^3} \Big[C_i(-q_b) G^{in}(q_b) T_{nm; k}(q_b - q_a) A_k(q_b - q_a) G^{ml}(q_a) F_{l;j}(q_a) A_j(q_a) \nonumber \\
    &\hspace{3cm}+ A_i(-q_b) F_{n;i}(-q_b) G^{nm}(q_b) T_{ml;k}(q_b - q_a) A_k(q_b - q_a) G^{lj}(q_a) C_j(q_a) \Big].
\end{align}
Substituting in the vertex functions of Eq.~\eqref{eq.VertexFuncs} and simplifying, we get
\begin{align}
    S_{\text{eff}}^{(1)}[A^2] 
    =& \frac{\tau_B^2}{\tau}\frac{\nu^3 e^2\alpha^2\omega_c^2 \epsilon^{il} \epsilon^{jk}}{32\pi^3\hbar^2 m v^2} \int \frac{d^3q}{(2\pi)^3}\, 
    (i\omega)^2 A_i(-q) \Big(G^{xx}(0) G^{lk}(q) + G^{xk}(q) G^{lx}(0) \Big) A_j(q).
\end{align}
The associated current is, after some index manipulation,
\begin{align}
    J_a^{(1;1)}(\omega) =& -\frac{\tau_B^2}{\tau}\frac{i\nu^3 e^2\alpha^2\omega_c^2\omega \epsilon^{al} \epsilon^{jk}}{32\pi^3\hbar^2 m v^2} \Big(2G^{xx}(0) G^{lk}(\omega) + G^{xk}(\omega) G^{lx}(0) + G^{lx}(\omega) G^{kx}(0) \Big) E_j(\omega).
\end{align}
From this, the conductivity components can be obtained by evaluating the Levi-Civita elements and substituting in the relevant propagator terms, leading to longitudinal components
\begin{align}
    \sigma_{xx}^{(1;1)}(\omega) = -\nu\frac{e^2}{h}\frac{i\alpha^2\hbar\omega_c}{2\tau \Delta}\frac{\tau_B^2\omega_c\omega}{\omega_c^2 -\omega^2} \hspace{1cm} \text{and} \hspace{1cm} \sigma_{yy}^{(1;1)}(\omega) = -\nu\frac{e^2}{h}\frac{i\alpha^2\hbar\omega_c}{\tau \Delta}\frac{\tau_B^2\omega_c\omega}{\omega_c^2 -\omega^2}
\end{align}
and Hall components
\begin{align}
    \sigma_{xy}^{(1;1)}(\omega) = \nu\frac{e^2}{h}\frac{3\alpha^2\hbar\omega_c}{4\tau \Delta}\frac{\tau_B^2\omega^2}{\omega_c^2 -\omega^2} \hspace{1cm} \text{and} \hspace{1cm} \sigma_{yx}^{(1;1)}(\omega) = -\nu\frac{e^2}{h}\frac{3\alpha^2\hbar\omega_c}{4\tau \Delta}\frac{\tau_B^2\omega^2}{\omega_c^2 -\omega^2}.
\end{align}
The two Hall components remain antisymmetric, whereas the longitudinal
components are no longer equal, reflecting the symmetry breaking induced
by the tilt parameter $\alpha$. Since $\alpha\ll1$, these subleading
corrections are omitted from the main text and included here for
completeness. Up to this order in perturbation theory, the linear
conductivity is therefore
\begin{equation}
\sigma_{ab}^{(1)}(\omega)
\approx
\sigma_{ab}^{(1;0)}(\omega)
+
\sigma_{ab}^{(1;1)}(\omega).
\label{eq.SM_linear_conductivity_total}
\end{equation}

\section{Nonlinear conductivity and second-harmonic response}

The second-order conductivity follows from the cubic-in-\(A\) part of the first-order perturbative contribution to the effective action, \(S^{(1)}_{\rm eff}[A]\), given in Eq.~\eqref{eq.SM_Seff1}. This term provides the leading contribution to the bulk second-harmonic Hall response discussed in the main text. We therefore evaluate the \(\order{A^3}\) contribution to \(S^{(1)}_{\rm eff}[A]\), which yields the leading second-order current. This is represented diagrammatically by Fig. \ref{fig:diagrams}(f) and is given by
\begin{equation}
     S_{\text{eff}}^{(1)}[A^3] = \frac{i^2}{2\hbar^2}\int \frac{d^3q_a}{(2\pi)^3} \int \frac{d^3q_b}{(2\pi)^3} A_i(-q_b) F_{l_1;i}(-q_b) G_{l_1m_1}(q_b) T_{m_1l_2;j}(q_b-q_a)A_j(q_b-q_a) G_{l_2 m_2}(q_a) F_{m_2;k}(q_a) A_k(q_a),
\end{equation}
based on the Feynman rules. Substituting in the vertex functions from Eq.~\eqref{eq.VertexFuncs} and simplifying, including through manipulation of internal dummy index and momentum labels, the relevant action becomes
\begin{align}
    S_{\text{eff}}^{(1)}[A^3] =& \frac{i\tau_B^2\nu^3 e^3\alpha\omega_c}{32\pi^3 \hbar^2 m^2 v^3} \int \frac{d^3 q_a}{(2\pi)^3} \int \frac{d^3q_b}{(2\pi)^3} \, (\omega_b-\omega_a)\omega_a\omega_b \epsilon^{in} A_i(-q_b) G^{nx}(q_b) \epsilon^{jm} A_j(q_b-q_a) G^{ml}(q_a) \epsilon^{kl} A_k(q_a).
\end{align}
From this, the second-order current is given by
\begin{align}
    J_a^{(2;1)}(q) =& \frac{i\tau_B^2\nu^3 e^3\alpha\omega_c\omega}{32\pi^3 \hbar^2 m^2 v^3} \int \frac{d^3 q_a}{(2\pi)^3} \, \omega_a(\omega-\omega_a)\epsilon^{an}\Big(G^{nm}(q_a) G^{xl}(q-q_a) \\
    & + G^{nx}(q) G^{lm}(q_a) + G^{nl}(q) G^{xm}(q_a) \Big) \epsilon^{jm} A_j(q_a) \epsilon^{kl} A_k(q-q_a), \nonumber
\end{align}
or, in terms of the homogeneous electric fields,
\begin{align}
    J_a^{(2;1)}(\omega) =& -\frac{i\tau_B^2\nu^3 e^3\alpha\omega_c\omega\epsilon^{an}}{32\pi^3 \hbar^2 m^2 v^3} \int \frac{d\omega_a}{2\pi}\Big(G^{nm}(\omega_a) G^{xl}(\omega-\omega_a) \\
    & + G^{nx}(\omega) G^{lm}(\omega_a) + G^{nl}(\omega) G^{xm}(\omega_a) \Big) \epsilon^{jm} E_j(\omega_a) \epsilon^{kl} E_k(\omega-\omega_a), \nonumber
\end{align}
From this, we obtain the eight second-order conductivity tensor elements, describing the response
\begin{align}
    J_a^{(2)}(\omega)=\int\frac{d\omega_a}{2\pi} \sigma_{abc}(\omega_a,\omega-\omega_a)E_b(\omega_a)E_c(\omega-\omega_a),
\end{align}
which is permutation-symmetric, such that $\sigma_{abc}(\omega_a,\omega-\omega_a) = \sigma_{acb}(\omega-\omega_a,\omega_a)$, which is enforced by adding the two permutations and halving. Each of the eight elements can be written in the form
\begin{align}
    \sigma_{abc}^{(2;1)}(\omega_a,\omega-\omega_a) =& \nu\frac{e^3}{h}\frac{\alpha\hbar v}{8 \Delta^2 }\frac{\tau_B^2\mathcal{F}_{abc}(\omega_a,\omega-\omega_a)}{(\omega^2-\omega_c^2)(\omega_a^2-\omega_c^2)([\omega-\omega_a]^2-\omega_c^2)}, \label{eq.SM_sigma2_general}
\end{align}
with the polynomials $\mathcal{F}_{abc}(\omega_a,\omega-\omega_a)$ being
\begin{align}
    \mathcal{F}_{xxx}=&3\omega_c^2\omega(\omega^2\omega_a-\omega\omega_a^2)\\ \mathcal{F}_{xyy}=&\omega_c^2\omega(-3\omega^3+4\omega^2\omega_a-4\omega\omega_a^2+5\omega\omega_c^2)\nonumber\\
    \mathcal{F}_{yxx}=&i\omega_c\omega(3\omega^2 \omega_a^2 + 2\omega^2\omega_c^2-3\omega^3\omega_a-2\omega_c^4) \nonumber\\ \mathcal{F}_{yyy}=&i\omega_c^3\omega(4\omega^2-4\omega\omega_a+4\omega_a^2-6\omega_c^2)\nonumber\\
    \mathcal{F}_{xxy}=&i\omega_c\omega(3\omega^3\omega_a-6\omega^2\omega_a^2+3\omega\omega_a^3+2\omega^2\omega_c^2-4\omega\omega_a\omega_c^2+2\omega_a^2\omega_c^2-2\omega_c^4) \nonumber\\ 
     \mathcal{F}_{xyx}=&i\omega_c\omega(3\omega^2\omega_a^2-3\omega\omega_a^3+2\omega_a^2\omega_c^2-2\omega_c^4)\nonumber\\
    \mathcal{F}_{yxy}=&\omega_c^2\omega(4\omega^2\omega_a-4\omega\omega_a^2+3\omega_a^3-5\omega_a\omega_c^2) \nonumber\\ 
    \mathcal{F}_{yyx}=&\omega_c^2\omega(3\omega^3-5\omega^2\omega_a+5\omega\omega_a^2-3\omega_a^3-5\omega\omega_c^2+5\omega_a\omega_c^2).\nonumber
\end{align}
From this generalised formulation, we are interested in the case of second-harmonic generation, described by conductivity elements $\sigma_{abc}^{(2;1)}(\omega,\omega)$ yielding current $J_a^{(2)}(2\omega)$. In this case, we can write the elements in the form
\begin{align}
    \sigma_{abc}^{(2;1)}(\omega,\omega) =& \nu\frac{e^3}{h}\frac{\alpha\hbar v}{4 \Delta^2}\frac{\tau_B^2 \mathcal{F}_{abc}(X)}{(1-X^2)^2(1-4X^2)}, \label{eq.SM_sigma2_SHG}
\end{align}
in terms of the non-dimensionalised frequency: $X=\omega/\omega_c$, and where the polynomials are
\begin{align}\label{eq:Fabc_SHG}
    &\mathcal{F}_{xxx}(X)=-6X^4 \nonumber\\
    &\mathcal{F}_{xyy}(X)=-10X^2+16X^4 \nonumber\\
    &\mathcal{F}_{yxx}(X)=2iX-8iX^3 +12iX^5 \nonumber\\
    &\mathcal{F}_{yyy}(X)=6iX-12iX^3 \nonumber\\
    &\mathcal{F}_{xxy}(X)=\mathcal{F}_{xyx}(X)= 2iX-2iX^3-6iX^5 \nonumber\\
    &\mathcal{F}_{yxy}(X)=\mathcal{F}_{yyx}(X)=5X^2- 11X^4.  
\end{align}
In the main text, we also substitute $\tau_B^2\approx 1+\tau Y/2$, with $Y=\hbar\omega_c/\Delta\ll1$ being the ratio of the magnetic and band energy scales.
In the general second-order conductivity of Eq.~\eqref{eq.SM_sigma2_general}, we see three poles at each of the incident frequencies, $\omega_a$ and $\omega-\omega_a$, and the total frequency, $\omega$, matching the Kohn-mode frequency $\omega_c$, relating respectively to one-photon and two-photon resonances. In the second-harmonic conductivity of Eq.~\eqref{eq.SM_sigma2_SHG}, this leads to a second-order pole at $\omega=\omega_c$ due to the two incident frequencies being the same, as well as a first-order pole for the two-photon resonance at $2\omega=\omega_c$.
\section{Subleading loop corrections and convergence}

The leading nonlinear response presented in the main text is
obtained at tree level. Here, we examine loop contributions
arising at higher order in the tilt-dependent interaction.
As an example, we consider the loop correction to the
linear conductivity shown in Fig.~\ref{fig:diagrams}(e),
which originates from the second-order expansion of the
effective action in Eq.~\eqref{eq.SM_Seff2ndOrdPert}.
Since this diagram is connected, its contribution is
obtained from the first term of
Eq.~\eqref{eq.SM_Seff2ndOrdPert}, with the disconnected
contributions removed by the second term.
Introducing the prefactor
\begin{equation}
\mathcal{C}=\frac{\nu^2 e^2\alpha^2 v^2\hbar\omega_c^2}{128\pi^2\Delta^4},
\label{eq.SM_loop_prefactor}
\end{equation}
the connected second-order contribution is given by
\begin{align}
S^{(2)}_{\rm conn}[A]
=
i\hbar^2 \mathcal{C}\,
\frac{\epsilon^{\mu\beta}\epsilon^{\lambda\rho}}{Z_G[0,A]}
\int & d^3q_a\,d^3q_b\,d^3q_c\,d^3q_d\,
E_\mu(-q_a-q_b)\,
E_\lambda(-q_c-q_d)
\nonumber\\
&\times
\frac{\delta}{\delta j_\beta(-q_a)}
\frac{\delta}{\delta j_x(-q_b)}
\frac{\delta}{\delta j_\rho(-q_c)}
\frac{\delta}{\delta j_x(-q_d)}
Z_G[j,A]\bigg|_{j\to0} - S^{(2)}_{\rm disconn}[A]
\label{eq.SM_loop_expression}
\end{align}
From this connected-diagram contribution to the second-order effective action, the loop contribution is the term which has two propagators with two independent momenta, distinct from the higher-order tree-level contribution represented in Fig. \ref{fig:diagrams}(g). This loop term is given by the terms in the fourth functional derivative of $Z_G$ which yield two propagators and two delta functions in momentum which when integrated out causes the propagators to be connected at the two $E(q)$ terms. Additionally, loop terms are independent of $\mathcal{J}^{(0)}_i[A,q]$. From the Feynman rules, or equivalently by extracting
the loop terms from the connected contribution above, we obtain
\begin{align}
    S^{(2)}_{\text{loop}}[A] =& \frac{-i}{4\hbar^3} \int \frac{d^3q_a}{(2\pi)^3} \int \frac{d^3q_b}{(2\pi)^3} \Big[T_{nm;i}(q_b-q_a) A_i(q_b-q_a) G^{ml}(q_a) T_{lk;j}(q_a-q_b) A_j(q_a-q_b) G^{kn}(q_b) \Big],
\end{align}
which upon substitution of the vertex functions (Eq.~\eqref{eq.VertexFuncs}) and simplification of the resulting expression yields
\begin{align}
    S^{(2)}_{\text{loop}}[A] =& 
    -i\mathcal{C}\epsilon^{\mu\beta} \epsilon^{\lambda\rho}     
    \int \frac{d^3 q_a}{(2\pi)^3} \int \frac{d^3 q_b}{(2\pi)^3}\, \left[G^{xx}(q_b) G^{\beta\rho}(q_a) + G^{\beta x}(q_a) G^{x\rho}(q_b) \right] E_\mu(-q_a-q_b) E_\lambda(q_a+q_b).
\end{align}
Following some algebra, including relabelling dummy indices and momenta,
the corresponding loop correction to the physical linear current is 
\begin{align}
    J_{a}^{(1;2,\text{loop})}(q) =& \mathcal{C} \omega \epsilon^{a\beta} \epsilon^{\mu\rho} \int \frac{d^3 q_a}{(2\pi)^3} \big[ 2G^{xx}(q-q_a) G^{\beta\rho}(q_a) + G^{\beta x}(q_a) G^{x\rho}(q-q_a) + G^{x\rho}(q_a) G^{\beta x}(q-q_a) \big]  E_\mu(q),
\end{align}
The longitudinal conductivity components are

\begin{align}
\sigma_{xx}^{(1;2,\mathrm{loop})}(\omega)
&= 2C\omega \int \frac{d^3 q_a}{(2\pi)^3}
\left[
G^{xx}(q-q_a)G^{yy}(q_a)
+G^{yx}(q_a)G^{xy}(q-q_a)
\right],
\\
\sigma_{yy}^{(1;2,\mathrm{loop})}(\omega)
&= 4C\omega \int \frac{d^3 q_a}{(2\pi)^3}
\left[
G^{xx}(q_a)G^{xx}(q-q_a)
\right],
\end{align}
while the Hall components are 
\begin{align}
\sigma_{xy}^{(1;2,\mathrm{loop})}(\omega)
&= -4C\omega \int \frac{d^3 q_a}{(2\pi)^3}
\left[
G^{xx}(q-q_a)G^{yx}(q_a)
\right],
\\
\sigma_{yx}^{(1;2,\mathrm{loop})}(\omega)
&= -4C\omega \int \frac{d^3 q_a}{(2\pi)^3}
\left[
G^{xx}(q-q_a)G^{xy}(q_a)
\right].
\end{align}
Performing the angular and frequency integrations using the
Green's function in Eq.~\eqref{eq.SM_Gij_q}, we find that
the one-loop corrections to the Hall conductivity are
ultraviolet convergent. Similarly, power counting shows
that the one-loop correction to the second-order conductivity,
shown in Fig.~\ref{fig:diagrams}(h), is ultraviolet convergent.
In contrast, the longitudinal linear conductivity requires
ultraviolet regularisation. 
The corresponding momentum integration is therefore
restricted to the low-momentum regime
$\hbar k \ll p^\ast = 2mv$,
where the gradient expansion remains reliable.
This avoids the spurious softening of the collective-mode
dispersion obtained by extrapolating its truncated
polynomial form to large momenta.
Within this regime,
the longitudinal loop correction carries the perturbative
prefactor $\alpha^2/\Delta^4$, while the nonlinear
conductivity correction is parametrically suppressed
by an additional power of $\alpha/\Delta^2$,
carrying the prefactor $\alpha^3/\Delta^6$.
In any case, all these loop corrections are dynamical and vanish in the DC transport limit. 
\section{Finite-frequency nonlinear Hall resistivity}
We now obtain the nonlinear Hall resistivity without performing the
small-$X$ expansion used in the main text. Starting from Eq.~(2) of
the main text,
\begin{equation}
\rho^{(2)}_{TLL}(2\omega)
=
-\rho_{Ta}(2\omega)
\sigma^{(2)}_{abc}(2\omega)
\rho_{bL}(\omega)
\rho_{cL}(\omega),
\label{eq:rho2_full_def}
\end{equation}
we retain the unexpanded frequency dependence of the linear and
second-order conductivities within the large-gap and weak-tilt
approximation used throughout this work. 
For compactness, we recall $\tau_B^2 \approx  1+\frac{\tau Y}{2}$, $Y=\frac{\hbar\omega_c}{\Delta}$  and $X=\frac{\omega}{\omega_c}$. The linear resistivity tensor is obtained by inverting the conductivity tensor where its
determinant can be written as 
\begin{equation}
\det\boldsymbol{\sigma}(X)
=
\left(\frac{\nu e^2}{h}\right)^2
\frac{Q(X)}{1-X^2},
\qquad
Q(X)
=
1-X^2(1-\tau_B^2  )^2.
\label{eq:Qz_def}
\end{equation}
It is therefore convenient to introduce the dimensionless
longitudinal and Hall components
\begin{equation}
a(X)
=
\frac{i\tau_B^2 X}{Q(X)},
\qquad
b(X)
=-
\frac{1-(1-\tau_B^2 )X^2}{Q(X)}.
\label{eq:ab_def}
\end{equation}
The linear resistivity tensor then takes the form
\begin{equation}
\boldsymbol{\rho}(X)
=
\frac{h}{\nu e^2}
\begin{pmatrix}
a(X) & -b(X)\\
b(X) & a(X)
\end{pmatrix},
\label{eq:rho_linear_full}
\end{equation}
with the pole at $X=1$ not appearing in the resistivity due to cancellation of the $(1-X^2)$ factor when inverting $\sigma$.
Let the applied current make an angle $\theta$ with the crystalline
tilt axis, such that
\begin{equation}
\hat{\boldsymbol e}_L
=
(\cos\theta,\sin\theta),
\qquad
\hat{\boldsymbol e}_T
=
(-\sin\theta,\cos\theta).
\label{eq:LT_unit_vectors}
\end{equation}
The two incoming resistivity factors at frequency $\omega$ can be
collected into
\begin{align}
u_x(X,\theta)
=
a(X)\cos\theta-b(X)\sin\theta,
\qquad
u_y(X,\theta)
=
a(X)\sin\theta+b(X)\cos\theta,
\label{eq:u_full}
\end{align}
while the transverse projection of the outgoing resistivity tensor at
frequency $2\omega$ is described by
\begin{align}
v_x(X,\theta)
=
b(2X)\cos\theta-a(2X)\sin\theta,
\qquad
v_y(X,\theta)
=
b(2X)\sin\theta+a(2X)\cos\theta.
\label{eq:v_full}
\end{align}
Using the second-harmonic conductivity of Eq.~(24) of the main text,
the nonlinear Hall resistivity takes the closed form
\begin{equation}
\rho^{(2)}_{TLL}(2\omega,\theta)
=
-\frac{h^2}{\nu^2e^3}
\frac{\alpha\hbar v}{4\Delta^2}
\frac{\tau_B^2}
{(1-X^2)^2(1-4X^2)}
\,
\mathcal R(X,\theta),
\label{eq:rho2_full}
\end{equation}
where
\begin{align}
\mathcal R(X,\theta)
={}&
v_x
\left[
\mathcal{F}_{xxx}(X)u_x^2
+2\mathcal{F}_{xxy}(X)u_xu_y
+\mathcal{F}_{xyy}(X)u_y^2
\right]
\nonumber\\
&+
v_y
\left[
\mathcal{F}_{yxx}(X)u_x^2
+2\mathcal{F}_{yxy}(X)u_xu_y
+\mathcal{F}_{yyy}(X)u_y^2
\right].
\label{eq:R_full}
\end{align}
Here, the polynomial functions $\mathcal{F}_{abc}(X)$ are given in Eq.~\eqref{eq:Fabc_SHG}. 
Interestingly, just as in the linear resistivity, the poles corresponding to one- and two-photon resonances in the nonlinear conductivity are cancelled due to the tensor product with the three linear resistivities, with a factor of $(1-X^2)^2(1-4X^2)$ appearing in the numerator. Therefore, the remaining poles are $Y$-dependent, coming from the linear resistivity factors introducing additional denominators $Q(X)$ and $Q(2X)$. These $Y$-dependent poles occur at
\begin{equation}
X=\frac{1}{|\tau_B^2 -1|},
\qquad
X=\frac{1}{2|\tau_B^2 -1|},
\label{eq:rho_linear_poles}
\end{equation}
respectively. Since $\tau_B^2-1\simeq\tau Y/2$ in the large-gap limit,
these frequencies scale as $X\sim 2/Y$ and $X\sim 1/Y$ and therefore
lie outside the regime $\hbar\omega\ll\Delta$ in which the projected
low-energy theory is controlled.  
Equations~\eqref{eq:rho2_full} and \eqref{eq:R_full} retain the
unexpanded dependence on $X$ within the effective theory. Expanding
for $X\ll1$ and retaining terms through $\mathcal O(X^2)$ and linear
order in $Y$ gives
\begin{align}
\rho^{(2)}_{TLL}(2\omega,\theta)
\simeq{}&
\frac{h^2}{\nu^2e^3}
\frac{\alpha\hbar v}{4\Delta^2}
\Big\{
X^2
(6+11\tau Y)\cos\theta
+iX(2+\tau Y)\sin\theta
\Big\} + \mathcal O(X^3,Y^2).
\label{eq:rho2_low_frequency}
\end{align}
This reproduces the linear-in-$X$ imaginary response and the
quadratic-in-$X$ real response, together with the angular dependence discussed in the main text. The corresponding
frequency and angular dependence is shown in Fig.~2 of the main text.

\bibliography{refs}